\documentclass[a4paper,fleqn, review]{cas-dc}

\usepackage[square,sort,comma,numbers]{natbib}
\setcitestyle{square}
  \usepackage{todonotes}
\DeclareUnicodeCharacter{2061}{}
\usepackage{tikz}
\usetikzlibrary{shapes.geometric, arrows}
\usepackage{balance}

\usepackage{amsmath}
\usepackage{subcaption} 

\usepackage{algpseudocode}
\usepackage{algorithm}
\usepackage{bm}
\algdef{SE}[VARIABLES]{Variables}{EndVariables}
   {\algorithmicvariables}
   {\algorithmicend\ \algorithmicvariables}
\algnewcommand{\algorithmicvariables}{\textbf{Variables}} 

\usepackage{caption}
\def\tsc#1{\csdef{#1}{\textsc{\lowercase{#1}}\xspace}}
\tsc{WGM}
\tsc{QE}

\begin{document}
\let\WriteBookmarks\relax
\def\floatpagepagefraction{1}
\def\textpagefraction{.001}

\shorttitle{Rendering Transparency to Ranking in Educational Assessment via Bayesian Comparative Judgement}


\title[mode = title]{Rendering Transparency to Ranking in Educational Assessment via Bayesian Comparative Judgement} 




%

\ifdefined\DOUBLEBLIND
    \shortauthors{Anonymous \textit{et al.}}    
    \author[]{-- Author Names Removed for Peer-Review --}
\else
    \shortauthors{Gray \textit{et al.}}

\author[1,2]{Andy Gray}[type=editor,
    orcid=0000-0002-1150-2052,
    twitter=codingWithAndy,
]

\cormark[1]


\ead{a.gray2@bathspa.ac.uk}
\ead{445348@swansea.ac.uk}



\affiliation[1]{organization={Bath Spa University},
            city={Bath},
            country={United Kingdom}}
\affiliation[2]{organization={Swansea University},
            city={Swansea},
            country={United Kingdom}}

\author[2]{Alma Rahat}[
    orcid=0000-0002-5023-1371,
    twitter=AlmaRahat,
    ]


\ead{a.a.m.rahat@swansea.ac.uk}









\author[3]{Stephen Lindsay}[orcid=0000-0001-6063-3676]
\affiliation[3]{organization={University of Glasgow},
            city={Glasgow},
            country={United Kingdom}}

\ead{stephen.lindsay@glasgow.ac.uk}

\author[2]{Jen Pearson}[]


\ead{J.Pearson@swansea.ac.uk}

\author[2]{Tom Crick}[
    orcid=0000-0001-5196-9389,
    twitter=ProfTomCrick,]


\ead{thomas.crick@swansea.ac.uk}




\fi

\begin{abstract}
    Ensuring transparency in educational assessment has become a critical concern in contemporary discourse — perhaps more so post-pandemic — driven by the demand for more equitable and reliable evaluation methods. Comparative Judgement (CJ) has emerged as a promising alternative to traditional rubric-based assessments. However, the perceived opacity of decision-making processes within CJ has raised concerns about fairness and accountability. In this paper, our goal is to examine how the application of recently proposed Bayesian methods (by analysing data to update prior knowledge with observed data) to CJ (BCJ) can address these concerns and whether it truly offers a structured, and data-driven approach, that enhances transparency for both practitioners and learners: both qualitatively and quantitatively.

The BCJ introduces a probabilistic framework that incorporates prior information into the judgement process, allowing for more informed and interpretable decisions. By assigning probabilities to judgement outcomes, the Bayesian model provides a quantifiable measure of uncertainty, offering insights into the confidence associated with each decision. This enhanced transparency provides greater insights into the judgement process and the ucnertainty therein, facilitating a deeper understanding of how individual judgements contribute to the final ranking outcomes. BCJ systematically tracks how prior information, and successive judgements inform final decisions, thereby promoting greater transparency and accountability, with the possibility of identifying the level of disagreements between the assessors.

Additionally, a multi-criteria BCJ extends the standard BCJ approach by simultaneously addressing multiple learning outcomes (LO), mirroring the rubric-based marking structure. In this approach we use
an independent BCJ model 
per LO, resulting in more granular and transparent rankings aligned with specific goals. Importantly, it preserves the richness of information captured during CJ by offering detailed insights into each outcome. Furthermore, it facilitates the creation of an overall holistic ranking, drived from components models 
for individual rankings, offering a comprehensive evaluation of student work without compromising the depth of feedback that can be provided for each outcome. This approach enhances transparency by making both the criterion-specific and overall judgements more explicit, supporting fairer, more accountable assessments.

Using a real dataset arising from the delivery of a course in the higher education sector in the UK with a group of professional markers, in this paper, we demonstrate how the BCJ approaches can be applied in assessment contexts in a rigorous quantitative manner. We also show how the models can provide greater insight into the rationale of the ultimate ranking and behind individual and aggregate judgements. Additionally, we discuss how this approach facilitates external validation, ensuring that assessment practices remain not only transparent but also robust and defensible. Then, for the first time, through a semi-structured discussion with the participant markers and expert practioners in CJ, and qualitative analyses, we establish how useful it really is in identifying the rank, particularly in situations where transparency and auditability is paramount (for example, through “high stakes” national assessments), highlighting the benefits and drawbacks of adopting this in real-world academic practice, across a range of educational settings and contexts.

\end{abstract}





\maketitle

\section{Introduction}
    \label{sec:intro}

    The transparency of assessment practices in education is a significant concern, particularly in light of recent global shifts that have underscored the need for fairer, more rigorous, and accountable assessment systems \cite{Crick2021}. In the UK, these challenges are exacerbated by the heavy workload pressures faced by teachers, with assessment being one of the primary contributors \cite{Morris2023Can, working_lives_teachers_leaders_2024}. Time pressures on marking can lead to inconsistencies, making fairness and transparency in educational assessment paramount \cite{Rasooli2018Re-conceptualizing, Tierney2014Fairness}.

    Teachers in England work an average of 54 hours per week, while school leaders work more than 60, according to the Department for Education (DfE)’s workload survey. Excessive time spent on unnecessary tasks driven by an accountability regime contributes to the ongoing recruitment and retention crisis without improving pupil learning outcomes (LOs) \cite{neu_workload_advice}. Recognising this, the UK’s Secretary of State for Education has emphasised the need to reduce school workload, and the DfE has provided an online toolkit with practical resources to support school leaders and teachers in this effort \cite{school_workload_reduction_toolkit}.

   In 2019, the average self-reported working hours for all teachers and middle leaders was 49.5 hours per week, a reduction of 4.9 hours from 2016. Primary teachers and middle leaders reported working an average of 50 hours per week in 2019, down from 55.5 in 2016, while secondary teachers and middle leaders reported a decrease from 53.5 hours to 49.1 \cite{teacher_workload_survey_2019}. However, primary teachers continue to work longer hours than their secondary counterparts, though the gap has narrowed from 2 hours per week in 2016 to 0.9 hours in 2019. Clearly, there is a drive to make working conditions better in the UK, and arguably it has had some positive impact. 

    Nonetheless, Marking remains one of the most time-consuming tasks. According to the DfE’s workload survey, 61\% of secondary school teachers and middle leaders reported that they spend too much time marking \cite{working_lives_teachers_leaders_2024}. The proportion of teachers who feel overwhelmed by marking has remained persistently high, with 43\% reporting excessive marking workloads in 2024, compared to 46\% in both 2022 and 2023 \cite{working_lives_teachers_leaders_2024}.

    This significant impact of marking on teacher workload
    leads to stress, poor wellbeing, and inconsistencies in grading \cite{Jerrim2021When}. More importantly, high marking loads 
    exacerbates the element of subjectivity and bias in manual grading, particularly in large student populations with high teacher-to-student ratios \cite{Senanayake2024Rubric}. Furthermore, there is a disconnect between grading policy and practice, as many teachers do not consistently use written criteria, often relying on holistic rather than analytical judgments. This lack of standardisation leads to variability between different schools \cite{Bloxham2011Mark, Hausdorff1965The}.

    Traditional methods of evaluation, like rubric-based grading, although systematic, fail to address subjective inconsistencies and may not adequately capture the subtleties of student performance \cite{VelascoMartinez2019}. Comparative Judgement (CJ) has emerged as an alternative method with potential to overcome some of these limitations, offering a ranking system based on direct -- pairwise -- comparisons of student works rather than pre-defined scoring criteria \cite{jones2024comparative}. However, CJ may seem like a ``black-box approach" as assessors’ decisions and their underlying rationale are not transparent to students, educators, or other stakeholders \cite{Holmes2020}. To address these concerns, in this paper, we explore the application of Bayesian CJ (BCJ) to render the CJ process more transparent and interpretable.
    
    BCJ leverages a probabilistic framework that incorporates prior information into the CJ process, allowing a structured update of beliefs based on observed data \cite{GRAY2024100245}. This method is aligned with Bayesian principles, in which prior knowledge about assessments is updated as new judgements are made, thus refining the confidence of the model in ranking the results. By assigning probabilities to the outcomes of each judgement, BCJ not only quantifies uncertainty but also introduces a level of interpretability that was previously unattainable with traditional CJ because educators can see the degree of confidence associated with each ranking decision, promoting a clearer understanding of how judgements contribute to assessment outcomes. BCJ’s systematic tracking of prior information and successive judgements also provides an audit trail that enhances transparency, making it possible to trace back, and understand the sources of any discrepancies or \textit{measure} assessor disagreements.
    
    This transparent process has the ability to substantially improve assessment as it allows students to understand how their work is evaluated, cultivating trust in the fairness of assessments \cite{GRAY2024100245}. Moreover, BCJ offers assessors a data-driven basis for their ranking decisions, thus supporting consistency and fairness. By visualising confidence levels and quantifying uncertainty, the BCJ model highlights instances where judgements are not aligned \cite{gray2025bayesianactivelearningmulticriteria}. Such insights can help educational institutions implement measures to reduce biases and improve the reliability of assessments. Furthermore, BCJ enables external review and validation, as third parties can examine the probabilistic logic underpinning each judgement decision. This potential for external scrutiny reinforces accountability and adds a layer of credibility to the assessment process \cite{GRAY2024100245}.
    
    The multi-criteria Bayesian approach (MBCJ) integrates multiple LOs into the CJ framework \cite{gray2025bayesianactivelearningmulticriteria}. Traditional CJ operates primarily on a holistic judgement basis, often collapsing distinct LOs into a holistic decision of which of the pair is the overall best, and BCJ provides a Bayesian frameowrk for this. By contrast, MBCJ allows each LO to be assessed independently, enabling the derivation of independent rankings aligned with each specific LO \cite{gray2025bayesianactivelearningmulticriteria}. This nuanced approach provides an avenue for assessment of individual aspects, such as critical thinking, technical proficiency, and creativity, resulting in a more granular, detailed, assessment that mirrors rubric-based methods while maintaining the benefit of pairwise comparison structure of CJ. The multi-criteria BCJ method also provides an overall holistic ranking that combines the insights from each LO-specific model, offering a balanced view of a student's performance across various assessment criteria, much like the standard rubric-based approach \cite{gray2025bayesianactivelearningmulticriteria}.
    
    The MBCJ approach does not sacrifice depth for breadth; it retains the richness of qualitative insights inherent in CJ and expands them further. For example, if critical thinking and technical accuracy are weighted differently, the model can transparently reflect these nuances in the final ranking. This ability to disaggregate and re-aggregate rankings according to distinct LOs is particularly advantageous in high-stakes settings, where fair and transparent evaluation of multiple competencies is crucial, such as undergraduate final-year project dissertations. Through this process, MBCJ enhances transparency by clarifying the distinct and cumulative contributions of each LO to the final assessment, making the assessment process fairer and more comprehensible for all stakeholders \cite{gray2025bayesianactivelearningmulticriteria}.
    
    For the first time, to illustrate the effectiveness of BCJ in real-world educational contexts, this paper employs a dataset from a UK higher education course to demonstrate the practical implementation of BCJ and MBCJ, and evaluate their impact on transparency, fairness, and accountability in assessment. By employing a combination of quantitative analyses and qualitative insights from professional markers, and views of experts in the field, we assess the efficacy of BCJ and MBCJ in providing a clear rationale for individual and aggregate judgements. Furthermore, we examine the usefulness of BCJ and MBCJ for identifying sources of ambiguity or conflict in assessments, particularly in contexts that demand high levels of transparency, such as national assessments \cite{Holmes2020}. 

    In this study, the main contributions are:
    \begin{itemize}
        \item Illustration of the improvement of transparency of CJ rendered by BCJ as we show how it addresses criticisms of traditional CJ methods by offering a structured process that tracks and explains decision-making, and provides estimations of uncertainty in rankings. 
        \item Evaluation and comparison of traditional, BCJ and MBCJ -- an extension to BCJ from singular holistic comparisons to multi-criteria comparisons -- to standard marking within a real-world assessment context. The findings provide practical insights into which method performs better and when.
        \item Analyses of insights from educators’ experiences – through a combination of quantitative data analyses and discussions with professional markers and experts in CJ, the study explores how educators perceive these approaches in terms of fairness, workload, and usefulness. This ensures that the research is grounded in the realities of teaching and assessment.
    \end{itemize}

    We review the literature and background in section \ref{sec:lit_background}. Section \ref{sec3} looks at the experiment's setup and the methodologies used. Section \ref{sec:res} looks at the results and discusses them, while in section \ref{sec:con}, we make our conclusions about the study.

\section{Literature and Background}
    \label{sec:lit_background}

    In section \ref{subsec:ass}, we explore the importance of assessments within education and the traditional approach to markings concerns. In section \ref{subsec:cj}, we review the literature around the rapidly growing approach to assessments called CJ. In section \ref{subsec:bcj}, we explain the process of BCJ, and in section \ref{subsec:MBCJ}, we explore the multi-criteria approach to BCJ, or MBCJ.

    \subsection{Assessment}
        \label{subsec:ass}
    
        Traditional marking where we assign a value or score to a piece of work under assessment
        is the dominant form of grading in education. In this method, teachers assign marks based on fixed criteria or rubrics, aiming to gauge a student's performance in an absolute sense. Despite its widespread use, this approach has been critiqued for issues related to consistency, bias, transparency, and, crucially, the cognitive demands it places on educators.
    
        On inconsistency, even when a grading rubric is in place, researchers have shown that different teachers can interpret the same criteria in varying ways, leading to discrepancies in scoring \cite{DeMoira2002Marking, Scharaschkin2000The}. Additionally, biases, such as those based on a student's previous performance, personality, or even handwriting, can unintentionally affect marks. \cite{Scharaschkin2000The} highlights the ``halo effect," where a teacher’s perception of a student's past achievements influences their grading of current work. This effect can lead to an overestimation or underestimation of a student’s true capabilities, which is particularly problematic in high-stakes assessments. 
    
        Biases can also stem from factors like teacher fatigue, stress, and subjective preferences. Researchers have shown that teachers’ grading decisions are often influenced by non-academic factors, even subconsciously. For example, teachers may give higher marks to work that aligns more closely with their own views or personal standards \cite{Willey2010Improving}. These biases undermine the fairness of assessments, which can affect students' opportunities and their trust in the educational system \cite{Guskey2024Addressing}.

        It should be noted that while anonymisation is often deemed as a definitive way to combat such biases and improve trusts in the system, evidence suggests that it may not necessarily be effective (see, for example, \cite{pitt2018impact}). Also, it is not always feasible to use anonymisation for all types of assessments, for instance, for presentation.

        Transparency in assessment refers to the clarity and openness with which grading criteria and decisions are communicated to students and other stakeholders. Often, traditional marking lacks this transparency, as students may receive little feedback beyond an overall numerical score or grade. This can leave students unclear about what specific aspects of their work need improvement. A lack of transparency in absolute marking hinders students’ learning processes, as they cannot fully understand how they are being evaluated or how they might improve in the future \cite{Bamber2015The}.
    
        Furthermore, when grading is inconsistent or biased, students and parents may feel frustrated or sceptical about the fairness of the assessment process. This lack of transparency is especially concerning in high-stakes situations, where grades can significantly impact future educational and career opportunities. Transparency is vital to ensuring trust in the assessment process, yet traditional marking methods often fall short in providing the necessary clarity \cite{Ilahi2024Enhancing}.

        Traditional marking is time-consuming and cognitively demanding for teachers. Cognitive load theory suggests that teachers’ mental resources can be depleted by the effort required to assess large volumes of work accurately and consistently. This depletion is even more pronounced when teachers are required to mark open-ended or complex tasks, as these assessments require continuous decision-making and interpretation \cite{Pollitt2012}.
    
        The high cognitive load of marking often leads to fatigue, which in turn can reduce the accuracy and consistency of grades over time. Teachers under pressure to meet grading deadlines may resort to shortcuts, such as relying more heavily on first impressions or previously formed opinions about students’ abilities \cite{bramley2015investigating}. These shortcuts, while understandable, further compromise the reliability and validity of assessments. Additionally, time constraints and workload pressures can lead to ``marking fatigue", where teachers' grading quality deteriorates as they progress through a stack of assessments \cite{Grissom2015Strategic}.
    
        In terms of actual time spent, teachers would often spend hours outside of class reviewing and grading work in traditional marking \cite{Morris2023Can}. This not only impacts their workload but may also limit the time they have available for other important teaching activities, such as lesson planning and providing one-on-one support to students \cite{Jerrim2021When}. 
  
        It is clear that traditional absolute marking in education presents several challenges, including inconsistencies, bias, a lack of transparency, and the cognitive and time demands placed on teachers \cite{chen-et-al:2023}. These issues highlight the need for more reliable, transparent, and efficient assessment methods that support both educators and students in the learning process.

    \subsection{Comparative Judgement}
        \label{subsec:cj}

        Comparative judgement, or CJ for short, has emerged as a promising approach in educational assessment, offering an alternative to traditional scoring methods for evaluating complex, subjective tasks. Rather than scoring individual pieces of work with a numerical grade or rubric, CJ relies on the collective judgments of experts who perform pairwise comparisons of 
        student work and decide which of the two better represents the overall LO. CJ has demonstrated advantages in areas like reducing subjective bias and achieving higher reliability in assessments, which, as we discussed thus far, 
        are challenging to maintain with traditional methods \cite{Pollitt2012}.

        The conceptual roots of CJ date back to Thurstone's Law of CJ \cite{Thurstone1927}, which posits that humans are more consistent in making comparative rather than absolute judgments. In educational assessment, this model has gained popularity because it capitalises on the relative strengths of judges, enabling more accurate evaluations without relying on predefined criteria or rubrics \cite{bramley2015investigating}. By comparing two pieces of work at a time, assessors can avoid the common pitfalls of point-based marking, such as over-focusing on specific criteria, and certain biases.

        CJ as a structured method in education was popularised by Pollitt \cite{Pollitt2012}, who introduced a software platform allowing assessors to make pairwise comparisons systematically. This software enables the aggregation of judgments to produce a ranking order, where each student's work is placed relative to others based on the majority of comparisons made by judges. This method offers an innovative, user-friendly interface for assessors and facilitates faster and more reliable marking of complex work.

\begin{figure}
            \centering
            \includegraphics[width=1\linewidth]{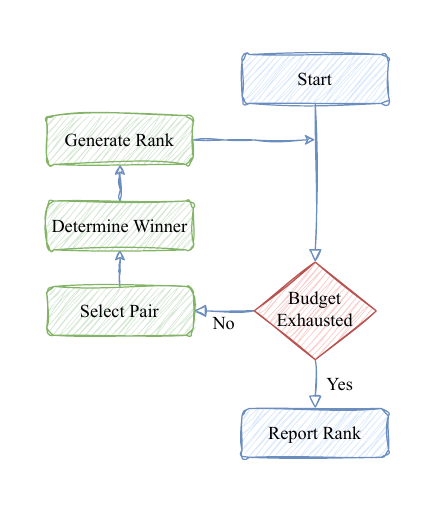}
            \caption{A flow chart depicting the CJ process. We start with a number of items to rank. Then, based on the budget on how many pairs we can show the assessors, we firstly select a pair to show. Then the assessor would pick the winner, and the statistical method in place would generate a rank for all the items in the light of new evidence. Once the budget is exhausted, we would report the final rank to the assessment owner. The green boxes are the core elements of CJ that varies methodologically between distinct approaches.}
            \label{fig:CJ-flow-chart}
        \end{figure}

        The overall CJ process is summarised in Figure \ref{fig:CJ-flow-chart}. We would start with all the items that are to be rank, and then select a pair of items that must be shown to the assessors for determining the winner. We would then deploy an appropriate statistical method to determine the complete rank, which would be reported at the end of the process, once we go beyond a specified threshold for the number of pairs to be shown to the assessors. 
        
        Different variants of CJ, considered in this paper, primarily vary in how the pairs are selected, how the winners are determined and how the overall ranks are produced. In standard CJ, the most popular approach towards selection of pairs is to select one uniformly randomly from all possible pairs \cite{jones2024comparative}. While it is possible that an automated process (e.g. a large language model \cite{gu2024survey}) to determine the winner in a paired comparison, in our context we typically observe that one or more humans make this decision synchronously or asynchronously as appropriate. Once the winner is known and the current dataset is augmented wiht this new information, we may use a statistical method to derive the overall rank. In traditional CJ, we use Bradley-Terry model (BTM) to generate the overall rank. In BTM, it is assumed that there is a likelihood over the score of an individual item, and then some approach towards maximising the likelihoods across all items, e.g. minorisation-maximisation method \cite{hunter2004mm}, to locate the optimal parameter set is used. The corresponding score to the optimal parameter set can be used to rank the items, all the way from $1$ to $N$, where the highest score is ranked $1$, and so on. 

        CJ has several advantages over traditional assessment methods. First, it has been shown to increase inter-rater reliability, meaning different assessors are more likely to agree on the ranking of work, even without detailed rubrics. For instance, \cite{bramley2019effect} found that CJ produced more reliable assessments for open-ended tasks, such as essays and projects, where defining a standard scoring rubric is challenging.
    
        Another benefit of CJ is its ability to reduce the time required for assessments. Instead of creating and training assessors on detailed rubrics, CJ relies on quick comparisons. According to studies, assessors often find it easier to determine which of two pieces of work is better than to evaluate individual work against a complex rubric. \cite{jones2015problem} highlighted this benefit and  reported that teachers assessing creative writing tasks could reach consensus using CJ quicker than with traditional methods.
    
        Moreover, CJ has been found to reduce biases that can creep into traditional scoring, such as the ``halo effect", where assessors’ impressions of one part of a work influence their judgment of other parts. Since assessors only view pairs in isolation, they are less likely to let factors unrelated to quality, such as familiarity with a topic or presentation style, sway their judgment \cite{leech2023holistic}.
    
    
        While CJ offers clear advantages, some critics argue that it lacks transparency compared to traditional grading \cite{ofqual2017}. One concern is that CJ does not provide detailed feedback on specific aspects of a student's work, as assessors are not evaluating against explicit criteria. As a result, students and teachers may find it harder to pinpoint areas for improvement \cite{Stuulen2024Effects}. This is a crucial gap that researchers have addressed by proposing MBCJ \cite{gray2025bayesianactivelearningmulticriteria}, and we demonstrate the efficacy of it through practical experimentations in this paper. 
    
        Another challenge lies in the scale of CJ tasks. While CJ works efficiently with small-to-medium-sized samples, the number of pairs that must be compared to decipher a reasonable ranking, grows exponentially due to combinatorial explosion. 
        This leads to practical limitations in time and resources \cite{jones2015problem}. Hence, users often pick a rule a thumb for the number of pairs that must be compared, for example, a common rule is to use ten times the number of items \cite{Pollitt2012}. In addition, software and methodological advances continue to streamline the process, making CJ more feasible for broader applications. For example, in BCJ and MBCJ, this is addressed via uncertainty-driven pair selection methods, enabling data collection in an optimal manner: in other words, the outcome ranking is the best one that can be derived given the amount of data that has been collected.
    
    
        In recent years, CJ has attracted attention as a tool for assessing complex competencies, such as critical thinking and creativity, which are difficult to measure with standard tests. Researchers are also exploring automated and semi-automated CJ systems, which could reduce the burden on human assessors and enable more widespread use \cite{bramley2019effect}.
    
        As we alluded thus far, Bayesian approaches are another area of growing interest within CJ research, as they allow for adaptive testing methods that can optimise the number of comparisons required by prioritising pairs that will yield the most useful information about relative performance \cite{GRAY2024100245}. This approach has shown promise in improving the efficiency and accuracy of CJ assessments, particularly in education settings where resources are limited. Beyond education, it has attracted new attention in other fields, e.g. in machine learning research for comparing performances of models with BTM based CJ \cite{wainer2023bayesian}. 
    
    
        CJ offers a compelling alternative to traditional assessment methods, especially for evaluating complex, open-ended tasks. Although it poses some challenges in terms of feedback and scalability, the method has demonstrated clear benefits in terms of reliability, efficiency, and fairness. As research and technology in this area advance, CJ is increasingly becoming a valuable tool in the landscape of educational assessment, potentially reshaping how teachers, students, and policymakers view assessment practices.

    \subsection{Bayesian Comparative Judgement}
        \label{subsec:bcj}
    
        Bayesian CJ (BCJ) is a novel approach to assessment that leverages Bayesian statistical machine learning methods to improve the reliability and transparency of CJ methods. 
        
        Firstly, we introduce intuitively how the Bayesian statistics work; interested readers should consult 
        \cite{lambert2018student}'s book for an accessible treatment of the topic. It all starts by defining a model for a process under scrutiny. The model should have at least one parameter that control the behaviour of the model, but can have many parameters that can be changed. We would then impose, based on experience or knowledge of the system, some prior probability density over the parameter(s). As we collect data, we would update the belief in the light of evidence, and produce an informed conclusion, i.e. known as the posterior density over the parameter(s). the more data we collect, the posterior density should become more accurate, and the uncertainty should collapse. At any point, the posterior is our best guess given the data, and we use that to make a judgement about the process model parameters, and consequently about the process outcome. 

\begin{figure}
            \centering
            \includegraphics[width=1\linewidth]{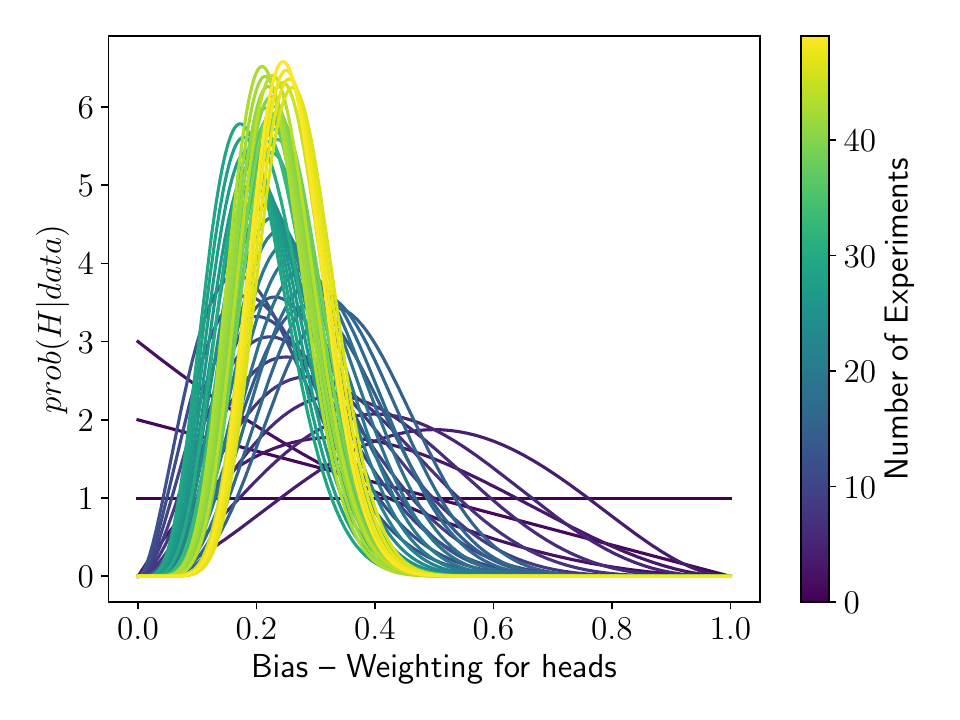}
            \caption{An illustration of Bayesian updates for a biased coin. The model is a generator for coin flipping outcomes: it will produce heads with the probability specified in the bias. Here, the bias is 0.3 (or 30\% chance of observing heads) and we collected data for 50 (simulated) coin flips and updated the prior belief to track the posterior density. Without any observations, the horizontal line at 1 depicts the flat prior belief that the bias could be anything. As we collect more data, the posterior density -- a Beta distribution for the Bernoulli bias variable -- over the bias narrows, i.e. gets confident about the estimation, with a mode around 0.3; the lighter colours are later estimates of the density. The illustration was inspired from the work of \cite{sivia2006data}.}
            \label{fig:bayes-coin}
        \end{figure}

        This is best appreciated through an example of a biased coin. The model we can assume is one which produces either heads (e.g. success/win) or tails (e.g. failure/loss) given a certain probability, and can be controlled by a bias parameter that controls the probability of observing heads. For example, if the bias is 0.3\%, that means we would observe heads around 30\% of the flips, and the rest would be tails. This type of biased-outcome parameter is often termed as a Bernoulli variable, and the posterior is known to be a Beta density which can be readily updated via the Bayes' theorem. As we collect more data, the posterior density becomes more confident, i.e. the uncertainty in the density reduces, and the most likely value starts to converge to the true bias. See Figure \ref{fig:bayes-coin} for an example for 50 coin flips for a biased coin with weighting 0.3.

        Now, the standard CJ (via BTM), estimate the expected individual scores for each item. These scores may not be straightforward to understand without training in statistics. Furthermore, there is little around uncertainty estimation, which is addressed by Bayesian versions of BTM in work done by \cite{wainer2023bayesian} that presents uncertainty around the scores. The key issues here the scores and the ranks derived from them are given to the assessors but the uncertainties about the ranks and the winners and losers in pairwise comparisons are not typically shown, or acknowledged. As a consequence, the assessors may derive undue confidence on the outcomes.
        
\begin{figure}
            \centering
            \includegraphics[width=1\linewidth]{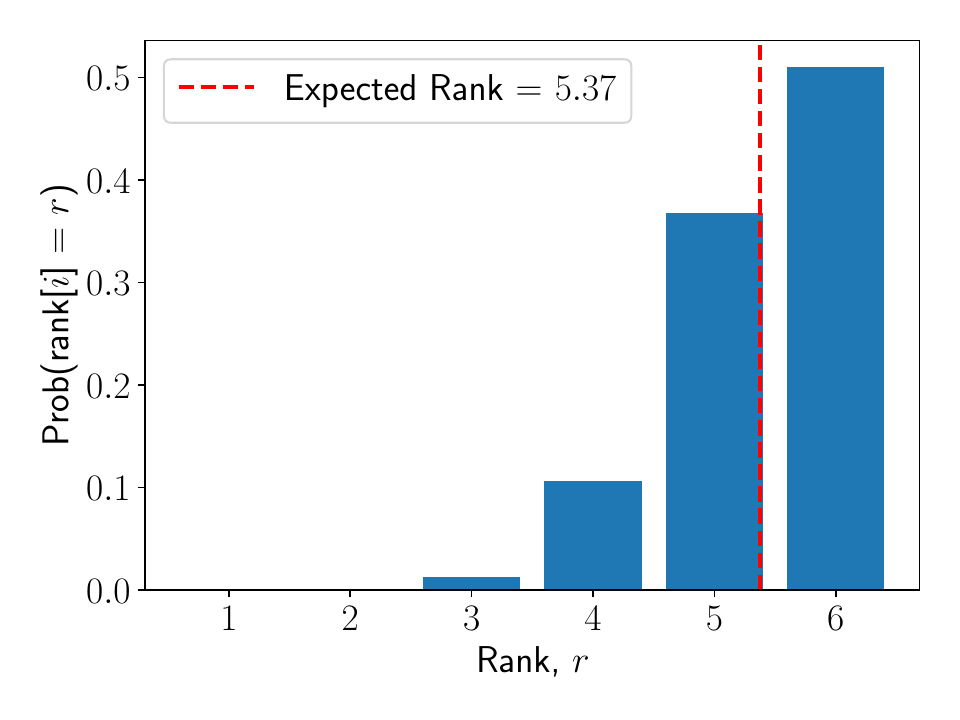}
            \caption{An example of rank density for an item $i$ post BCJ, given $6$ items. Here, this item has the highest probability (of around 50\%) of being ranked $6$, but the average (or expected) rank (shown in red dashed vertical line) is around 5.37 due to the consideration of uncertainty arising from paucity of data. BCJ uses the expected ranks (instead of scores in CJ) to determine the final ranks. }
            \label{fig:rank-density}
        \end{figure}

        \cite{GRAY2024100245} proposed an alternative Bayesian approach to CJ, which they call BCJ. In this approach, for generating ranks, the pairwise comparison is directly modelled as controlled by a Bernoulli variable or, in other words, a bias towards an item over another -- effectively making pairwise comparison analogous to coin flipping. This allows them to capture the uncertainty in comparison. As a result, this renders greater clarity in how much judges are agreeing on a per-pair basis. They then show how to derive probabilistic interpretations of ranks for individual items, accumulating the posteriors across all possible pairs assuming independence between pairs for comparison outcomes; we provide an illustration of rank density in Figure \ref{fig:rank-density}. This helps in deciphering how the rank densities are distinguishing the items, and the expected ranks five us a way to decipher an overall rank of the items. For selection, they proposed, for the first time, a way to use entropy -- that is proportional to the width shown in Figure \ref{fig:bayes-coin} -- as a measure for identifying the most uncertain pair, and therefore the most informative pair to show the judges next. They demonstrated via synthetic and real experiments that BCJ, combined with the active learning approach, outperforms traditional CJ methods in synthetic experiments with greater accuracy in ranking items with limited data.

        One of the most significant benefits of BCJ is its \textit{interaction} efficiency. \cite{GRAY2024100245} have shown that BCJ can achieve greater levels of accuracy as opposed to traditional CJ, especially when there is little data to make inference from, and thus providing an avenue to reduce the overall burden on assessors.
    
        Additionally, in BCJ, the uncertainty estimations allow assessors to identify pairs that are dividing the crowed, i.e. an estimated bias scoring close to 0.5 due to roughly similar number of wins and losses, and make a decision on which one is the clear winner. This feature allows for more nuanced decisions in cases where traditional methods might struggle to distinguish between similar levels of performance \cite{GRAY2024100245}. By acknowledging these subtleties, BCJ can produce more accurate and trustworthy results, which is especially useful for complex assessments such as written essays or project-based work.

        The use of CJ in educational assessment is growing, particularly for subjects and skills that are difficult to measure with standardised tests. For example, critical thinking, creativity, and communication skills are all competencies that do not lend themselves well to fixed-answer formats but can be effectively evaluated using comparative judgment \cite{Pollitt2012}. BCJ’s efficiency and adaptability make it feasible for applications in education, where resources and time are often constrained \cite{GRAY2024100245}.

        Despite its advantages, BCJ is not without challenges. One concern is that BCJ systems can be complex to set up and require computational resources to calculate Bayesian probabilities in real-time. While this has not been a barrier in experimental studies with smaller number of items to compare, it could pose challenges for schools or smaller educational institutions with limited technology support. That is why a Monte-Carlo (MC) version of the BCJ approach was also presented by \cite{GRAY2024100245}.

        Overall, BCJ represents an exciting advancement in educational assessment, combining the reliability of CJ with the efficiency of Bayesian methods. While there are challenges to its widespread adoption, BCJ has the potential to improve both the effectiveness and efficiency of assessments for complex, open-ended tasks. As research continues, BCJ could become an increasingly valuable tool for educators, helping to ensure fairer and more accurate assessments in diverse learning environments.

    \subsection{Multi-criteria Bayesian Comparative Judgement}
        \label{subsec:MBCJ}

        As discussed before, CJ has been criticised for not considering multiple dimensions of comparisons: this reduces the richness of information for both students and assessors. Despite the accuracy conferred by the approach, this is an important shortcoming. BCJ also suffers from this. Hence, traditional marking has a clear advantage in this aspect, because judges consider the different areas of the rubric with overall grade derived as a weighted sum of components, and thus complete information is available and can be queried.

\begin{figure}
            \centering
            \includegraphics[width=7.5cm]{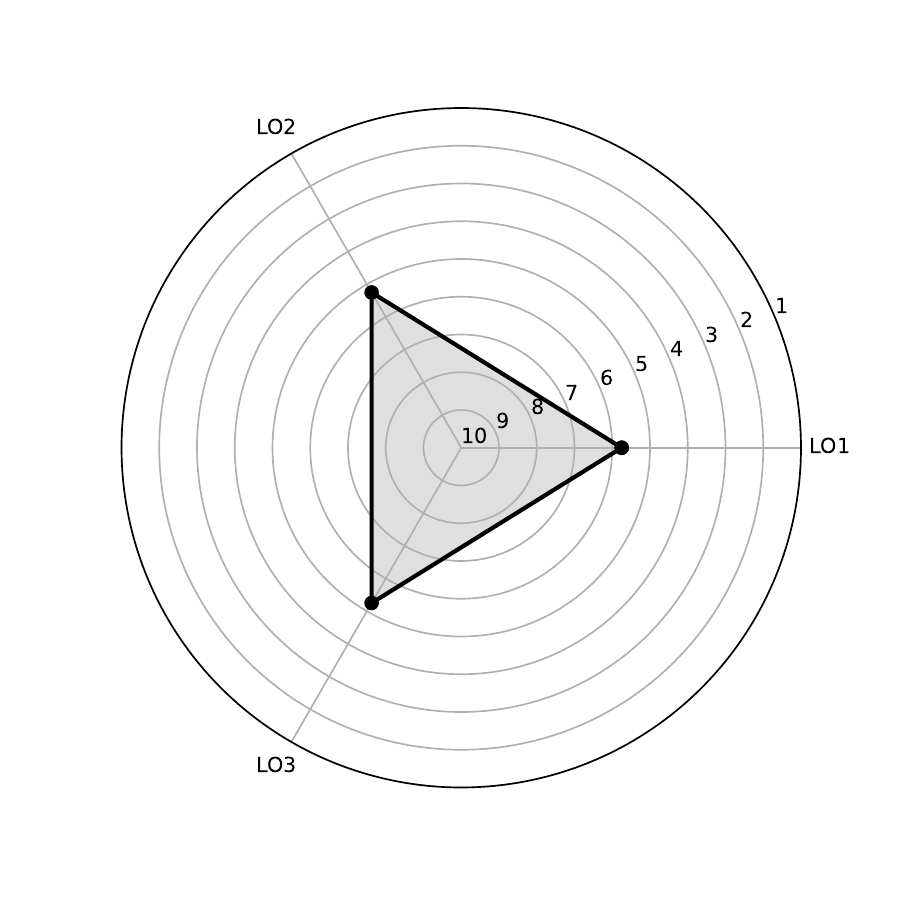}
            \caption{A radar plot depicting an item's expected rank $\mathbb{E}[r]$ ($5.75, 5.25, 5.25$) performance across three different LOs derived from component wise paired comparisons between 10 items. While conferring the same level of transparency for overall rankings as BCJ, this provides a detailed look into the components and how an individual item performs across the LOs. This helps educators to identify areas where the candidate would possibly need personalised intervention.}
            \label{fig:radar_plot}
        \end{figure}

        \cite{gray2025bayesianactivelearningmulticriteria} explored the idea of creating a multi-criteria version of BCJ for the first time. In their approach, pairwise comparison is performed in each individual component (or criterion) within a rubric, but at a time across all components for a pair that is being evaluated. This data allows component specific estimation of ranks per item and the probability densities therein. They then show how to combine them via weighted sum of the component cumulative densities, and produce an overall rank density. Like BCJ, the expected ranks of items can be used to derive overall rankings. In addition, they also proposed an extension to the pair selection method based on combined entropy -- again an estimation of the maximum utility of the next pair to be shown to the participant assessors. An illustration of the outcome expected ranks from the process is shown in Figure \ref{fig:radar_plot}. Their experiments show that MBCJ is equivalent to BCJ in performance, without the loss of rich component wise comparisons. In addition, they proposed a novel reliability measure for both BCJ and MBCJ. 

        For full mathematical details of BCJ and MBCJ, readers should refer to the original papers: \cite{GRAY2024100245} and \cite{gray2025bayesianactivelearningmulticriteria} respectively.

        In this paper, our goal is to evaluate BCJ and MBCJ in real world context with real markers and experts in CJ, and see how they fare against absolute marking. We start this discourse with a description of the experimental setup in the next section.

\section{Experimental Settings}
    \label{sec3}

This research considers various factors to determine educators' opinions on conventional grading, standard BCJ, and multi-criteria BCJ. Coursework submissions are evaluated, and marks supplied by the module's lead lecturer serve as a ground-truth score for comparison. Three marking assistants drawn from the module perform traditional absolute marking, standard BCJ, and MBCJ. Due to their role within the university, all have expertise in delivering and marking the module's assessments. The process is discussed in more detail in section \ref{subsec:research_approach}.

Once the marking assistants had completed all three marking approaches, they were asked to answer a questionnaire (see Appendix \ref{app:HCI_questions}) and, later, are brought together to discuss the approaches used in a workshop (see Appendix \ref{app:workshop}. The techniques used are explained in section \ref{subsec:research_approach}. We present the findings to industry experts who carry out research within academia on CJ in an educational setting and people working within industry who look at the policies around assessment for government and exam boards who also research into how CJ can be used in educational settings (see Appendix \ref{app:expert_int}), and present our findings later in the paper. 

\begin{figure*}[ht!]
    \includegraphics[width=\textwidth]{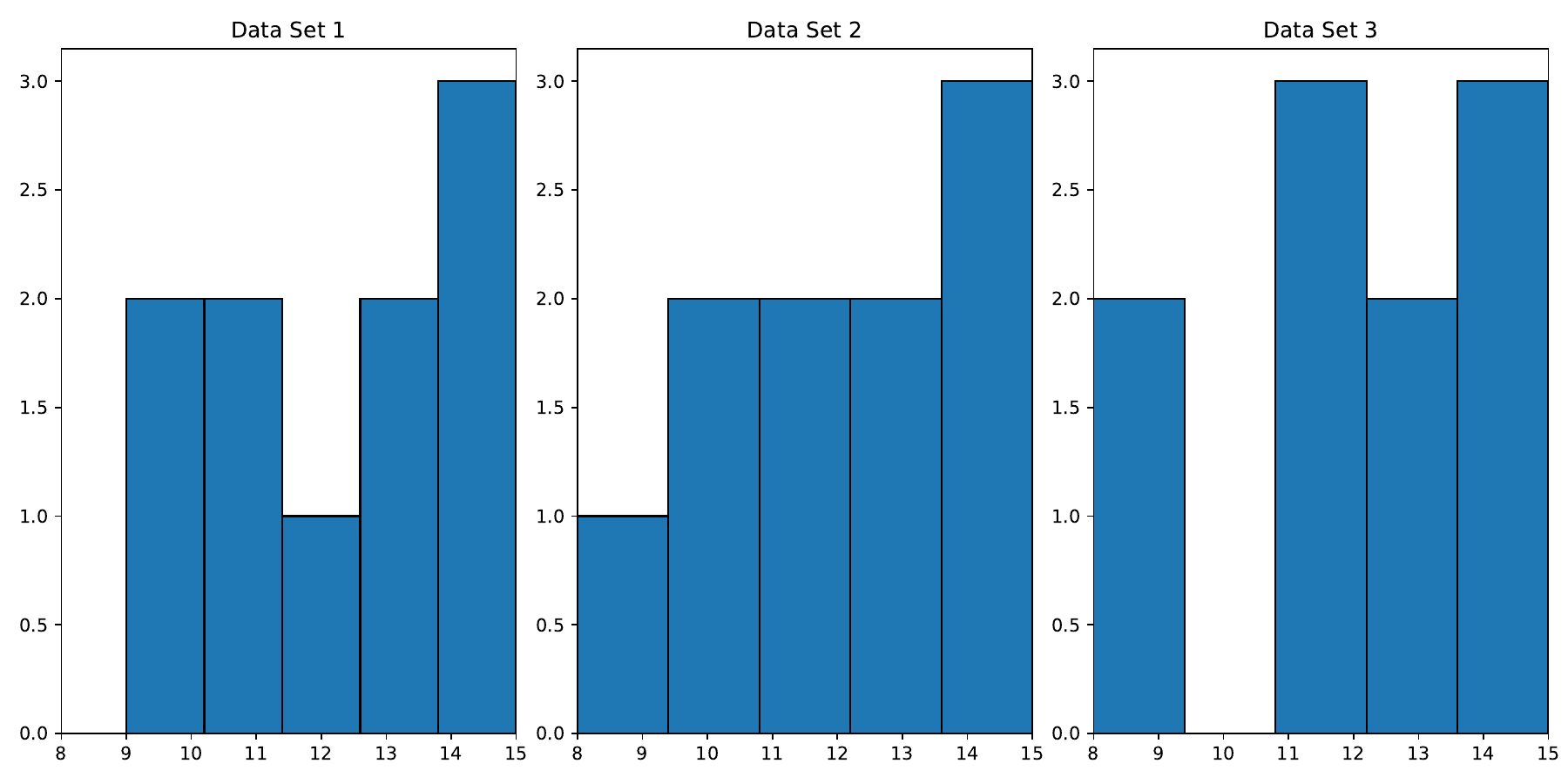}
  \caption{A histogram of marks for submissions in different groups: candidates for traditional marking, BCJ and MBCJ, entitled as dataset 1, 2 and 3 respectively. Clearly, the groups have similar distribution over the range between 8 and 15; this is important for a fair comparison between the groups.}
  \label{fig:grade-dist}
\end{figure*}

\subsection{Dataset}

We received from the lead lecturer marks for $30$ pieces of work that were submitted at a master's level course, where the prompt was to critically review a recent research paper. The submissions were anonymised with any identifying detail (e.g. student ID) removed: we received a distinct identifier with no relationship with the student and a corresponding mark. We created three groups, each with $10$ pieces, for traditional rubric based, BCJ and MBCJ marking, using a stratified sampling approach \cite{neyman1992two}. The distributions of the marks (out of 20) is given in Figure \ref{fig:grade-dist}.

The assessment criteria for the critical review included \textit{five} key areas: Introduction and summary, which requires a student to provide a clear explanation of the topic and a concise summary of the aims, main findings, and key arguments; Quality of Analysis and Evaluation, where the assessor looks for originality, a strong evaluation of the paper's strengths and weaknesses, and highlights of any unique aspects; Conclusions, which involves testing students for providing a summary of the perspectives presented and discussing any future impact or work; Writing quality, where the assessor would focus on organisation, structure, effective transitions, and grammatical accuracy of the submission; References, where students ability to provide relevant references and correct formatting are tested. It should be noted that the assignment brief that explained the criteria and the tasks were available to the markers for preparation prior to the experimentations.

    \subsection{Web Interface for Experimentation}
        \label{subsec:web-app}

            BCJ and MBCJ marking were done through web apps. The designs for the applications are similar with the main difference between the two apps being that the standard BCJ approach only had a single button for each item being displayed. In contrast, the multi-dimension application had a button for each LO that was being compared against, ensuring that only one button could be pressed for one of the items. 

\begin{figure}[h]
                \centering
                \includegraphics[width=\linewidth]{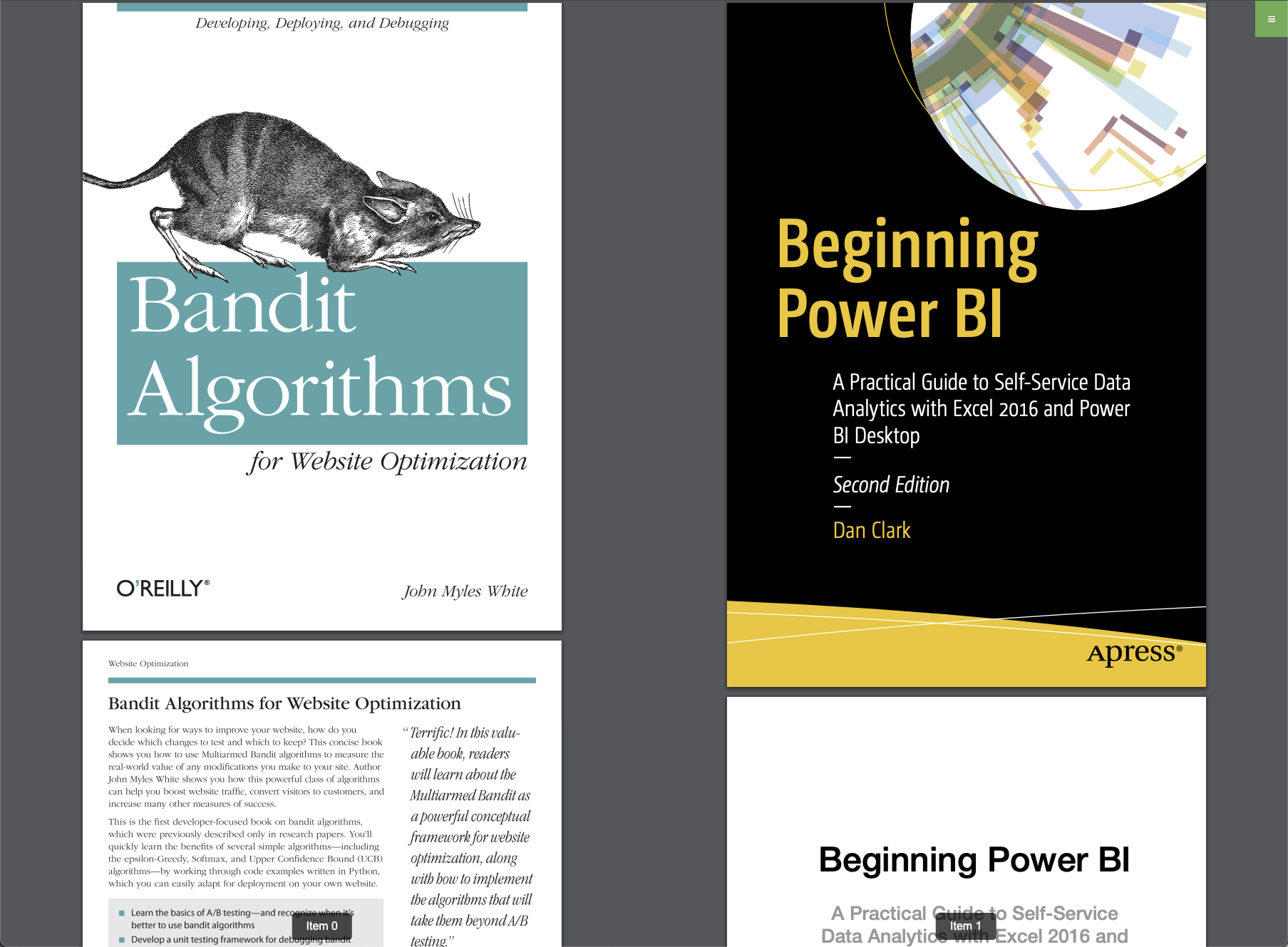}
                \caption{An example of the web app page for the standard BCJ's comparison. This page is what the assessor will see when they are making their judgements on the items being presented to them. Once they have pressed the corresponding button linked to the item they prefer, this will update the scores and then produce two new items for the assessor to compare.}
                \label{fig:sd-display}
            \end{figure}

            Figure \ref{fig:sd-display} shows the application's interface for the single-dimension version. The user is presented with two items that are being compared and two buttons. The user presses the button related to the item they deem to be of higher quality. 
            Once the button has been pressed, this updates the results matrix for the entropy picking method and then selects a new pair of items for the assessor to make a judgement on. Assessors can continue until they want to stop. However, it is recommended that a minimum of the number of items ($N$) times $10$ comparisons are completed \cite{jones2024comparative}.

\begin{figure}[h]
                \centering
                \includegraphics[width=\linewidth]{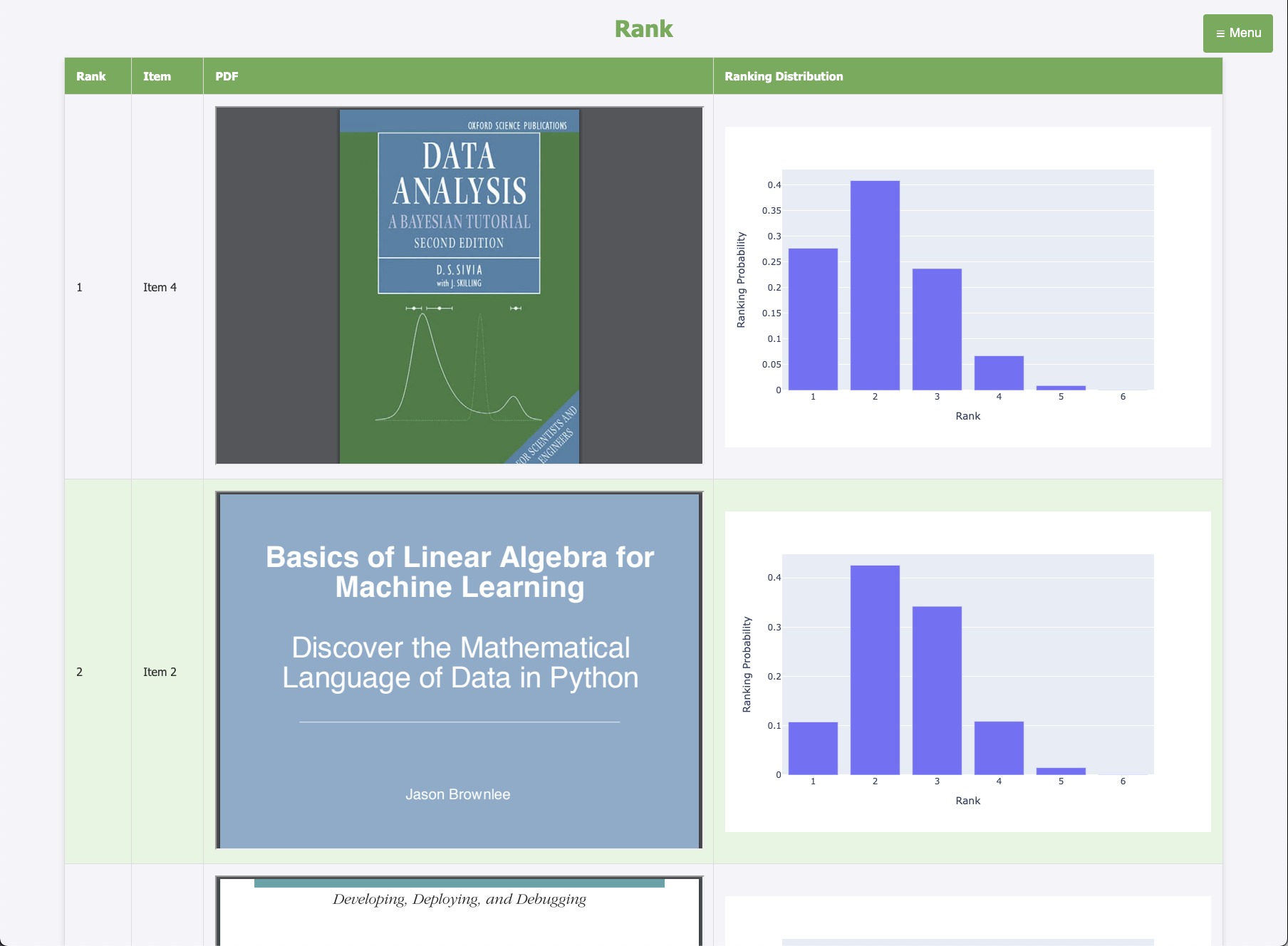}
                \caption{ When the assessor wants to view the results, they can visit the results page. This web app page shows the items in order of their ranking, so the item ranked first will appear first on the page, and as the assessor scrolls down the page, they will then view the additional items until they reach the last ranking item. Each item's rank, a copy of the item and their ranking distribution are shown to the assessor, ensuring maximum transparency is present to them on how the decisions have been made.
                }
                \label{fig:sd-results}
            \end{figure}

            When assessors want to view the ranking of the items, they can view the results page, as demonstrated in Figure \ref{fig:sd-results}. The items are rendered in rank order, so the highest ranked item appears first, and the weakest item, as they score it, will be at the bottom of the page. A graph is shown alongside the items depicting the ranking distributions of the items that have been compared. The ranks are calculated from the performance matrix using either BCJ or MBCJ while navigating to the results page. 

\begin{figure}[h]
                \centering
                \includegraphics[width=\linewidth]{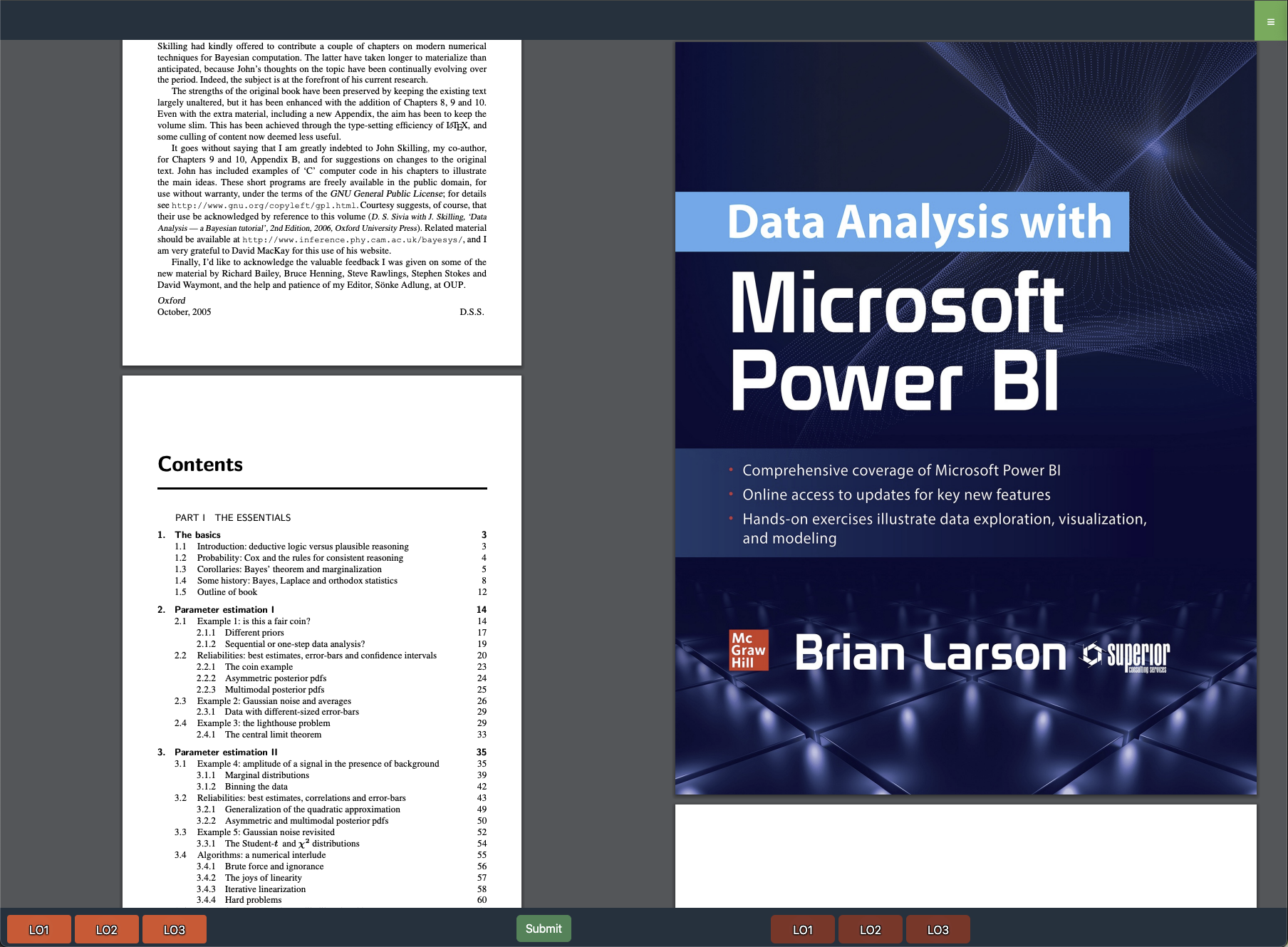}
                \caption{This figure shows the interface for the web apps page for the multi-criteria comparison page. Like the standard BCJ page, this is where the assessor will make their decisions on the items displayed. However, they will need to make decisions based on individual LOs this time. They press the submit button once they are ready to submit their preferences. This will update the LOs results and then produce two new items on which to make judgements.}
                \label{fig:md-display}
            \end{figure}

            Figure \ref{fig:md-display} shows the comparison screen that the assessors view when making their pair-wise comparisons. This screen is very similar to the standard approach (see Figure \ref{fig:sd-display}), but it has some key differences. One of the key differences is the items have a button for each LO that is being assessed as well as a submit button. When an assessor presses, for example, LO1 button for item A it will light up to show it has been selected, if they were to press LO1 for item 2, the button will  dim back to the default colour to ensure that only one LO for each item is selected. Once the assessor is happy with the selections, they hit the submit button and each LO's preference matrices are updated, which enables the differential beta entropy to select the next pair of items to present to the assessor.

\begin{figure}[h]
                \centering
                \includegraphics[width=\linewidth]{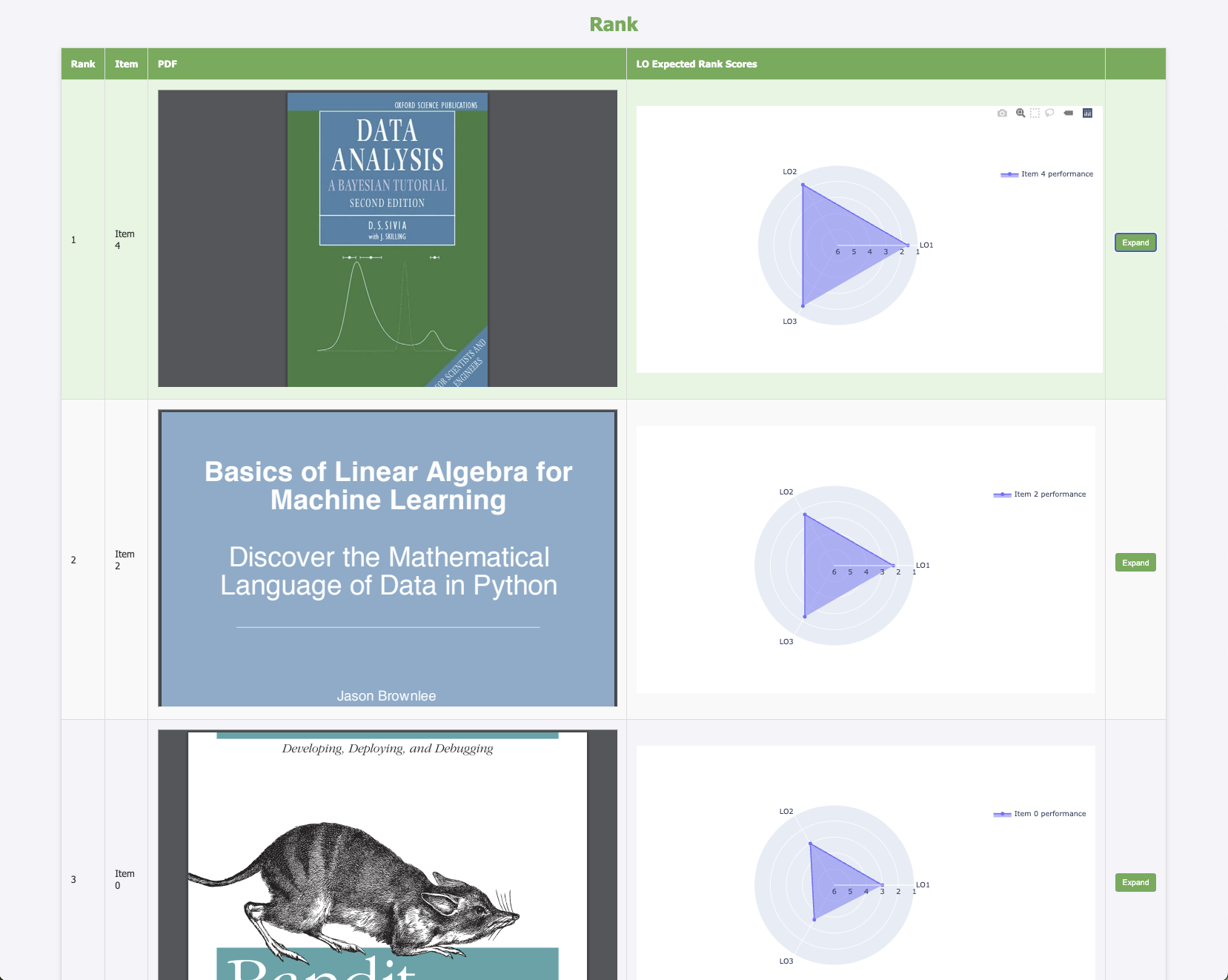}
                \caption{This figure presents the results of a multi-criteria BCJ web app showcasing transparency in ranking distributions across various LOs. The page provides an overview of the item's overall rank and and where the item's expected rank $E_r$ scores for each of the LOs in a radar plot. An expand button is available for the assessor to be able to view the ranking distributions for the overall score and individual LOs.}
                \label{fig:md-results}
            \end{figure}

            Figure \ref{fig:md-results} presents the results of a multi-criteria Bayesian CJ web app, showcasing transparency in ranking distributions across various LOs. The visualisation provides an overview of the item's overall rank and distribution and its performance within each LO.

            The results section shows the ranking distributions, which are multiple bar charts illustrating the frequency distribution of rankings for the item across all LOs. The overall ranking distribution is initially shown, similar to the standard BCJ web app, but there is an additional button that enables the user to expand or collapse a dropdown area that enables the ranking and distributions of the individual LOs for each item. Three additional bar charts represent the ranking distributions for each LO, allowing for easy comparison and analysis.

            The overall rank of the item is displayed alongside the item itself, providing a clear indication of its performance in relation to other items. Showing the distribution of rankings for the item across all LOs offers insight into its performance within each outcome as a holistic overall.

            The results section of the web app provides a clear and transparent representation of the item's ranking distributions across various LOs. The overall rank and performance metrics offer valuable insights for educators and researchers seeking to understand the item's strengths and weaknesses in relation to other items.

            \begin{figure}[h]
                \centering
                \includegraphics[width=\linewidth]{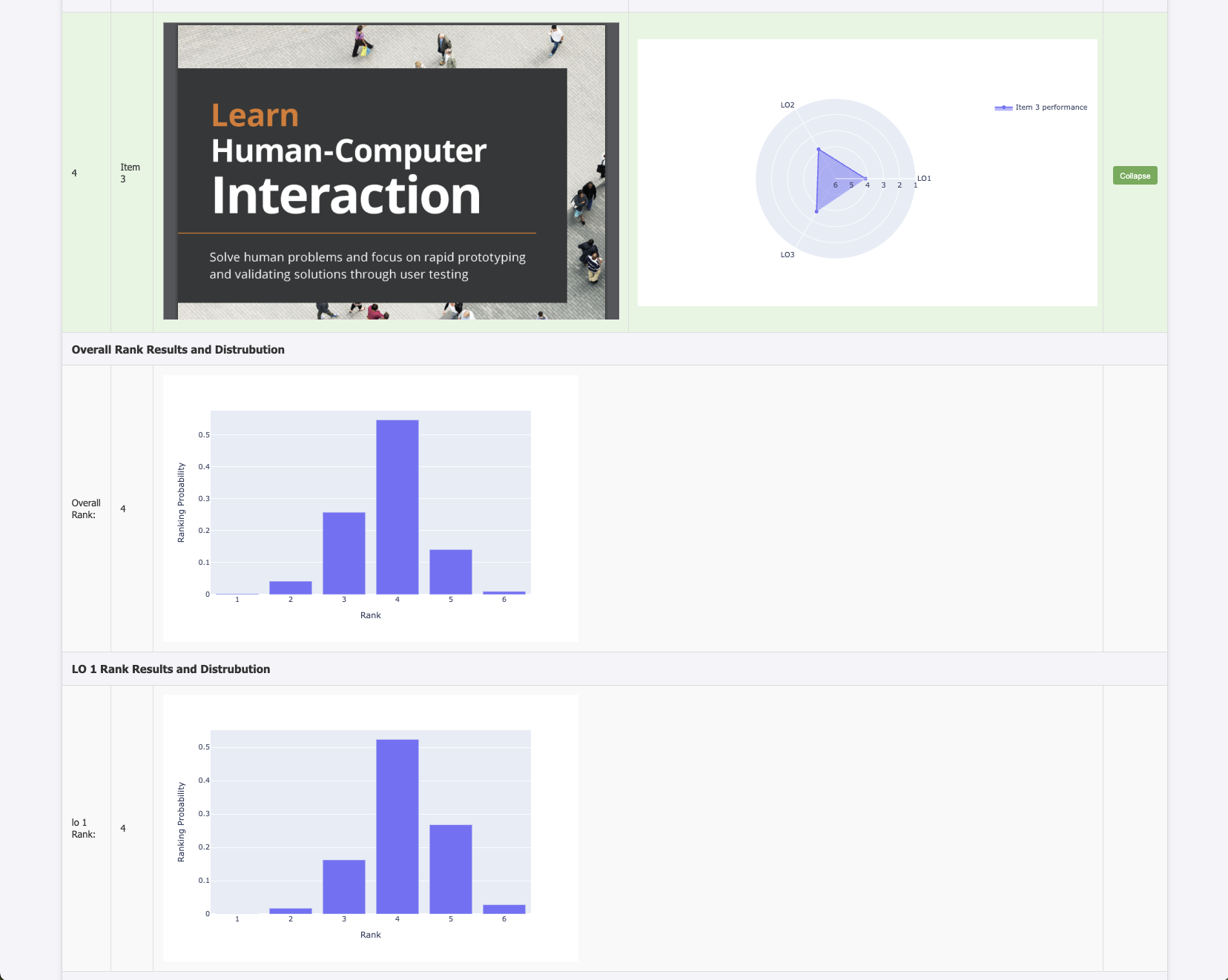}
                \caption{This figure presents the results of a multi-criteria BCJ web app showcasing transparency in ranking distributions across various LOs. The page provides an overview of the item's overall rank and distribution and its performance within each LO. }
                \label{fig:md-results-expanded}
            \end{figure}
            




    \subsection{Research Approach}
        \label{subsec:research_approach}

        Markers were given as much time as needed to complete the traditional marking, and a maximum of 2 hours to complete comparisons for each of the CJ methods. The markers did all three methods in different orders to try and mitigate against familiarising with the marking criteria over time. Once the markers had completed the three approaches, they took part in the semi-structured questionnaire (see Appendix \ref{app:HCI_questions}), done in isolation from the other markers. After this, all three markers come together to discuss their experiences as a whole  (see Appendix \ref{app:workshop} for an outline of the initial workshop plan).


    \subsection{Normalised $\tau$ Rank Distance Metric}
        The $\tau$ rank distance metric is used to see how the ranks generated by the individual assessors compare against an ``oracle" (the lead lecturer to the module's assessment) scores. The oracle's scores are used as the target rank to enable us to measure variation in the different marking methods against the oracles' official marks that were returned to the students. This does bring up other questions, especially ethical ones, like what methods are actually most accurate as marking is such a perceptual thing. However, these discussions are beyond the scope of this experiment.  
        
        We calculated $\tau$ distance via normalised Kendall's $\tau$ rank distance. The is calculated by counting the discrepancies between lists. The greater the distance, the more disparate the lists~\cite{kendall1938new, fagin2003comparing}. The normalised distance ranges from 0 (indicating perfect agreement between the two lists) to 1 (indicating complete disagreement between the lists). For example, a distance of $0.03$ means that only $3\%$ of the pairs differ in ordering. In this paper, when a method progressed, we noted the $\tau$ distance after each paired comparison, and this showed how well the relevant method converged to the target rank. This will represent the accuracy, and, therefore, the validity, of the ranks that come out of different methodologies.

\section{Results and Discussion}
    \label{sec:res}

    In this section we look at the $\tau$ results against the target ranks, the markers' questionnaires, and the outcomes of the workshop session that took place with all the markers as well as expert interviews. We focus on if and how the BCJ and MBCJ are transparent and the issues raised to be mindful of when using BCJ and MBCJ.

    \subsection{$\tau$ Results Against Target Rank}

        Marker one completed their traditional marking sample in two hours and 17 minutes (see Table \ref{tab:trad_results}). Compared to the Oracle's ranking, this produced a $\tau$ score of $0.3556$ , marker two 
        $0.4$ and  marker three $0.4$.

        \begin{table}[h]
            \centering
            \caption{A Table summarising the performance of markers during the traditional absolute marking process. The table includes key metrics such as the total time spent by each marker, the number of pairwise comparisons conducted, and the corresponding Tau scores. 
            }
            \label{tab:trad_results}
            \begin{tabular}{llll}
                \cline{1-3}
                Marker & Time    & Tau Score   \\ \cline{1-3}
                1      & 2:17:03 & 0.3556      \\ \cline{1-3}
                2      & 1:26:00 & 0.4           \\ \cline{1-3}
                3      & 1:54:00 & 0.4           \\ \cline{1-3}
                
            \end{tabular}
            
        \end{table}

        When we compare the markers with each other (see Table \ref{tab:vs_trad_results}), marker one and two generated a $\tau$ score of $0.4$, while marker one against marker three generated a $\tau$ score of $0.4$, marker two and marker three generated a $\tau$ score of $0.4888$. 
        
        These $\tau$ scores show that the markers were as far apart from each other in respect to  their final ranks as they were from the target rank of the oracle and, in the case of marker two and three, more so at $0.4888$. This suggests that variation is a result of noise, as opposed to bias, in the process.   

         \begin{table}[h]
        \centering
            \caption{The $\tau$ results of the final ranks created by the three markers when compared against each other for traditional marking. We can see that these compared to each other are not as close compared to the $\tau$ results from the oracle's rank in Table \ref{tab:trad_results}, but we can see that marker one compared to marker two and marker one compared to marker three were the most similar with marker two and three being the furthest away.}
            \label{tab:vs_trad_results}
            \begin{tabular}{lll}
                \cline{1-3}
                Marker & vs & Tau Score   \\ \cline{1-3}
                1      & 2  & 0.4        \\ \cline{1-3}
                1      & 3  & 0.4       \\ \cline{1-3}
                2      & 3  & 0.4888        \\ \cline{1-3}
            \end{tabular}
            
        \end{table}


        Performing BCJ, marker one completed $49$ comparisons in one hour and seven minutes. Their $\tau$ score was $0.1556$ (see Table  \ref{tab:bcj_results}), while marker two took two hours to complete their comparisons and completed a total of $46$ with a $\tau$ score of $0.1778$ and marker three took one hour and twenty-two minutes and done a total of $50$ comparisons and had a $\tau$ score of $0.1778$.

        These results show that all markers became closer to the oracle's target ranks doing BCJ. When comparing these $\tau$ results to the traditional marking methods, which were marked similarly to the oracle's, this shows that traditional methods are quite inconsistent and that the marks when marking in an absolute manner are very subjective and can vary from marker to marker. The BCJ approach removes many inconsistencies as it enables the markers to compare pair-wise, which has been identified in the literature as a more suitable approach to how the human brain works.

        \begin{table}[h]
        \centering
            \caption{A Table summarising the performance of markers during the BCJ process. The table includes key metrics such as the total time spent by each marker, the number of pairwise comparisons conducted, the resulting rank assigned based on their contributions, and the corresponding $\tau$ scores. 
            }
            \label{tab:bcj_results}
            \begin{tabular}{llll}
                \cline{1-4}
                Marker   & Time & No. of Comparisons & $\tau$ Score   \\ \cline{1-4}
                1        & 1:07:00 & 49              & 0.1556      \\ \cline{1-4}
                2        & 2:00:00 & 46              & 0.1778      \\ \cline{1-4}
                3        & 1:22:00 & 50              & 0.1778      \\ \cline{1-4} 
                Combined &         &                 & 0.1111       \\ \cline{1-4}
            \end{tabular}
            
        \end{table}

        When we compare the markers against each other (see Table  \ref{tab:vs_bcj_results}), marker one and two generated a $\tau$ score of $0.3333$, in contrast, marker one against marker three generated a $\tau$ score of $0.2889$, marker two and marker three generated a $\tau$ score of $0.2667$, these results show that the markers become more aligned with each other compared to traditional marking.

        While the markers are showing that they are more aligned with the oracle's ranking when it comes to BCJ, interestingly, while they are more in line with each other compared to traditional marking, they are further away from each other regarding $\tau$ scores than they are compared to the oracle's mark.

        \begin{table}[h]
        \centering
            \caption{The $\tau$ results of the final ranks created by the three markers when compared against each other for BCJ. We can see that these are not as close compared to each other as the $\tau$ results compared to the oracle's rank in table \ref{tab:bcj_results}, but we can see that marker two and three were the most similar, with marker one and three the next closest.}
            \label{tab:vs_bcj_results}
            \begin{tabular}{lll}
                \cline{1-3}
                Marker & vs & $\tau$ Score    \\ \cline{1-3}
                1      & 2  & 0.3333       \\ \cline{1-3}
                1      & 3  & 0.2889       \\ \cline{1-3}
                2      & 3  & 0.2667       \\ \cline{1-3}
            \end{tabular}
            
        \end{table}

        With MBCJ, marker one completed $37$ number of comparisons in two hours. Their $\tau$ score was $0.1333$ (see Table  \ref{tab:MBCJ_results}), while marker two took one hour and forty-nine minutes to complete their comparisons and done a total of $57$ with a $\tau$ score of $0.1556$ and marker three took one hour and twenty-two minutes and done a total of $53$ comparisons and had a $\tau$ score of $0.2886$.

        These results show that all markers were even more in line with the oracle's target ranks compared to traditional and BCJ for MBCJ, except for marker three whose $\tau$ score was $0.1108$ higher than their BCJ score but $0.1114$ lower than traditional marking. However, marker one was $0.0223$ better than BCJ and $0.2667$ than traditional. while marker two was $0.0222$ against BCJ and $0.2444$ against traditional.

        \begin{table}[h]
            \centering
            \caption{A Table summarising the performance of markers during the MBCJ process. The table includes key metrics such as the total time spent by each marker, the number of pairwise comparisons conducted, the resulting rank assigned based on their contributions, and the corresponding $\tau$ scores. 
            }
            \label{tab:MBCJ_results}
            \begin{tabular}{llll}
                \cline{1-4}
                Marker & Time & No. of Comparisons & $\tau$ Score  \\ \cline{1-4}
                1        &  2:00:00    & 37          & 0.1333      \\ \cline{1-4}
                2        &  1:49:00    & 57          & 0.1556      \\ \cline{1-4}
                3        &  1:22:00    & 53          & 0.2886      \\ \cline{1-4} 
                Combined &             &             & 0.2         \\ \cline{1-4} 
            \end{tabular}
            
        \end{table}

        When comparing the marker's final ranks against one another, we can see that markers one and two were most aligned with $0.1111$ and then markers two and three were the next most in line with a $\tau$ score of $0.1778$ while markers one and three were furthest apart with a score of $0.2444$.

        \begin{table}[b]
            \centering
            \caption{The $\tau$ results of the final ranks created by the three markers when compared against each other for multi-criteria BCJ.}
            \label{tab:vs_MBCJ_results}
            \begin{tabular}{lll}
                \cline{1-3}
                Marker & vs & $\tau$ Score   \\ \cline{1-3}
                1      &  2    & 0.1111   \\ \cline{1-3}
                1      &  3    & 0.2444   \\ \cline{1-3}
                2      &  3    & 0.1778   \\ \cline{1-3}
            \end{tabular}
        \end{table}

        Overall, we can see that the results from the traditional marking approach reveal noticeable differences in both the time taken and the consistency of rank ordering among the markers. Marker one took the longest to complete the task, spending two hours and 17 minutes, and obtained a $\tau$ score of $0.3556$, indicating a moderate level of alignment with the Oracle's ranking. In contrast, markers two and three completed their marking more quickly (one hour, twenty-six minutes, and one hour and fifty-four minutes, respectively) and achieved a slightly higher $\tau$ score of $0.4$. This suggests a marginally better agreement with the target rank compared to marker one, despite the variation in time spent. The comparative analysis between markers shows that marker two and marker three were the least aligned with each other, yielding a $\tau$ score of $0.4888$, highlighting greater inconsistency in rank ordering.

        The BCJ method demonstrated greater consistency between markers than traditional marking. Marker one completed $49$ comparisons in one hour and seven minutes with a $\tau$ score of $0.1556$, while marker two conducted 46 comparisons in two hours, achieving a $\tau$ score of $0.1778$. Marker three completed the highest number of comparisons ($50$) in one hour and twenty-two minutes, matching $\tau$ of marker two of $0.1778$. In particular, the $\tau$ scores for BCJ are lower overall compared to traditional marking, suggesting a closer alignment with the Oracle rank. Furthermore, the markers were more consistent with each other, as indicated by the lower $\tau$ scores compared to each other, with the closest alignment observed between markers two and three ($\tau$ score of $0.2667$).
        
        MBCJ further enhanced alignment with the Oracle's ranking. Marker one conducted $37$ comparisons in two hours with a $\tau$ score of $0.1333$, while marker two performed 57 comparisons in one hour and forty-nine minutes, obtaining a $\tau$ score of $0.1556$. Marker three completed $53$ comparisons in one hour and twenty-two minutes with a $\tau$ score of $0.2886$. These results indicate that MBCJ generally led to better alignment with the Oracle than BCJ and traditional marking, particularly for markers one and two. However, marker three's performance was less consistent, showing a higher $\tau$ score than in BCJ. When comparing markers against each other, the highest alignment was between marker one and two ($\tau$ score of $0.1111$), suggesting that MBCJ facilitated a more consistent approach to ranking.
        
        These findings indicate that traditional marking yielded the highest variability both in terms of time spent and rank consistency. In contrast, BCJ and MBCJ improved consistency across markers and alignment with the Oracle's ranking. MBCJ demonstrated the best overall performance, although individual differences among markers remained evident. This suggests that CJ methods, particularly MBCJ, could offer a more reliable alternative to traditional absolute marking by reducing subjective variability.




    \subsection{Questionnaire Results and Analysis}


        Marker one rated traditional marking as moderately easy to use (three), noting that well-defined criteria made it manageable but still cognitively demanding. Transparency was also rated a three, as the feedback process was clear at an individual level but lacked comparability between students. They were less confident in the accuracy of their marks (two), recognising the potential for inconsistency due to subjectivity and fatigue over time. They approached marking by evaluating each LO separately, weighting them accordingly, but did not particularly enjoy the process. A structured template for students was suggested as an improvement to streamline marking.

        BCJ was found to be more difficult than traditional marking, with them rating ease of use as two. They struggled with the holistic nature of the comparisons, as they typically assessed work LO by LO rather than as a whole. Transparency was also rated low (two), as the process lacked clear justifications for the rankings beyond the final distribution. Their confidence in the rankings was rated a three, as comparative ranking helped highlight relative quality but increased subjectivity. They found BCJ more cognitively demanding, especially early on, and would not recommend it over traditional marking.
        
        MBCJ was rated higher in ease of use (four), as LO-delineated comparisons aligned better with their marking approach. They found it significantly more transparent (four), appreciating the radar plot that visualised individual strengths and weaknesses. Their confidence in the rankings was also rated four, as the structured approach reduced subjectivity. Although initially cognitively demanding, they found it became easier over time while maintaining objectivity. They suggested adding an “unsure” option for cases where two submissions were indistinguishable.
        
        Marker one preferred MBCJ over other methods, as it aligned with how they assessed student work and provided clearer comparative insights. While traditional marking was familiar and felt “safe”, they believed MBCJ had the potential to improve consistency, particularly when multiple markers were involved. They were at their most confident in MBCJ’s rankings, as its structured approach reduced inconsistencies in subjective judgement. However, they noted that traditional marking still offered more direct feedback to students, which they felt could be integrated into MBCJ in the future.

        Marker two found that traditional marking was the easiest and most transparent method, rating both aspects a five. They appreciated the structured nature of the process, which allowed for clear criteria-based assessment. They noted that providing feedback enhances transparency but mentioned that an even more detailed mark scheme would be beneficial. However, they were somewhat uncertain about the accuracy of their marks, rating that between three and four. They expressed a preference for structured marking but acknowledged that issues such as inconsistent student presentation could impact the experience.

        For BCJ, the participant rated its ease of use a four, citing challenges in comparing papers of similar quality without standard criteria. Initially, they found the method to be exhausting, particularly since it was the first they attempted. Transparency was rated between four and five, as they appreciated the probability distributions but felt it remained somewhat "black boxy." They were fairly confident in the ranking results, but noted that their lack of understanding of the underlying algorithm reduced their confidence slightly. They preferred marking individual sections explicitly rather than making holistic judgements.
        
        Regarding MBCJ, marker two found it significantly easier than BCJ, rating it a five for ease of use. They appreciated the ability to compare work across LOs, which made it more transparent than BCJ. Confidence in the rankings was also rated highly, as they could see how individual components contributed to overall scores. However, they pointed out that the method lacked explicit feedback, which they viewed as essential for student improvement.
        
        When asked about their preferred method, they acknowledged that MBCJ was more efficient but favoured traditional marking due to its transparency and ability to provide feedback. They believed BCJ and MBCJ were useful but would work best alongside traditional marking rather than replacing it. Ultimately, they had the most confidence in the rankings generated by traditional marking, as it provided clear, section-by-section scores rather than relative comparisons between students.

        Marker three found traditional marking to be the most transparent but also the most time-intensive and mentally demanding. They rated its ease of use as a two, citing the need to apply specific criteria, distinguish between similar scores, and provide feedback. However, they rated transparency as a five, as traditional marking clearly breaks down the reasoning behind each score, though they acknowledged that consistency among markers is crucial. They felt the process was somewhat accurate (three to four) but prone to variability based on the marker’s mood or level of fatigue. While they found the approach familiar and structured, they did not enjoy it due to its time-consuming nature and the need to create extensive comments for student feedback.

        For BCJ, the participant found it significantly easier, rating it a four or five. They appreciated the simplicity of pairwise comparisons, particularly when differences between submissions were clear. However, they found transparency lacking (two to three), as it was difficult to pinpoint why a particular ranking emerged, especially over time. While seeing the rank distributions helped somewhat, they felt it would not provide enough actionable feedback for students. They were fairly confident (four) in the final rankings, as they aligned with their expectations, though they recognised that inconsistencies in marking could influence results. Compared to traditional marking, they found BCJ generally less mentally taxing, except when comparing closely matched submissions.

        Regarding MBCJ, the participant found it more balanced, rating ease of use between three and four. They liked its ability to break down performance across multiple LOs, which made transparency stronger (rated four). They felt this approach provided clearer insights into strengths and weaknesses across criteria. Confidence in the rankings was also high (four), as the method captured differences in individual components while maintaining overall consistency. However, they noted that minor variations, such as differences in referencing, could lead to occasional inconsistencies.

        When asked about preferences, the participant found BCJ to be the most straightforward but preferred MBCJ for its ability to highlight strengths and weaknesses across LOs. They believed MBCJ was a strong alternative to traditional marking, especially when multiple markers were involved, as it could help moderate inconsistencies. Ultimately, they had the most confidence in either traditional marking or MBCJ, with traditional marking being the safer, more familiar option but MBCJ offering potential advantages in efficiency and fairness. They suggested adding a flagging system to indicate particularly difficult comparisons or clear differences to refine the process further.






        The questionnaires found that traditional marking was associated with high levels of trust and transparency. However, MBCJ was perceived as offering greater transparency than BCJ, primarily because marking was conducted according to each LO. This allowed markers to clearly understand how judgements were made at a granular level, reinforcing their confidence in the method.

        MBCJ was generally preferred over BCJ, as markers felt it provided greater insight into the decision-making process. The structure of MBCJ aligned more closely with their usual marking practices, making it a more intuitive approach compared to BCJ. In contrast, BCJ was sometimes perceived as cognitively demanding, particularly when markers encountered two responses they judged to be of equal quality but lacked the ability to flag them as such. This forced them to engage in deeper reflection to make a final decision. Despite this, BCJ was still considered significantly less demanding than traditional marking and only marginally less so than MBCJ. Importantly, the slight increase in cognitive effort required for MBCJ was seen as a worthwhile trade-off, given its perceived transparency. Nevertheless, traditional marking remained the method in which markers placed the greatest trust, particularly regarding final marks and rankings.
        
        Both traditional marking and MBCJ were deemed more transparent than BCJ due to the ability to see how marks were assigned to individual LOs. Markers noted that if traditional marking required only an overall score rather than LO-based marking, its transparency would decrease, making the BCJ approach comparatively more acceptable. This highlights the significance of explicit marking criteria in fostering perceptions of fairness and clarity.
        
        Markers also acknowledged that the comparative nature of BCJ and MBCJ helped mitigate potential biases. In traditional marking, there is a risk that a marker may be overly harsh or lenient in their initial assessments before adjusting their expectations after encountering more responses. The CJ methods counteracted this by requiring markers to make direct comparisons between two pieces of work at a time, thereby reducing inconsistencies arising from fluctuating standards over the marking process.


        Across all three interviews, participants generally found traditional marking to be the most transparent but also the most time-consuming and cognitively demanding. While they rated its transparency highly, due to the structured nature of criteria-based assessment, they were less confident in its accuracy, citing concerns about subjectivity, inconsistency, and fatigue over time. They appreciated the ability to provide direct feedback to students but found the process mentally exhausting. Suggested improvements included providing students with structured templates to make marking more efficient and reduce ambiguity.

        BCJ was perceived as easier in some respects but introduced new challenges. While some found it straightforward when comparing submissions with clear quality differences, others struggled with its holistic nature, as it did not align with their typical LO-based marking approach. Transparency was rated lower than traditional marking, as the ranking process felt more like a ``black box" with limited justification for individual scores. Confidence in rankings varied, with some finding them reasonable but others feeling that the method increased subjectivity. Participants also found BCJ more mentally demanding than expected, especially when comparing closely matched submissions. One participant suggested incorporating an “unsure” option for cases where no clear distinction could be made between two pieces of work.

        MBCJ was consistently preferred over standard BCJ and, in some cases, over traditional marking. Participants found it more transparent and easier to use than BCJ, as breaking down comparisons by LO aligned better with their marking approach. They appreciated the radar plot visualisation, which provided clear insights into students’ strengths and weaknesses. Confidence in the rankings was higher than in BCJ, as participants felt that evaluating individual components led to more reliable outcomes. While still cognitively demanding, MBCJ was seen as fairer and more structured. However, they noted that it lacked direct feedback, which they considered essential for students’ learning.
        
        Overall, participants preferred MBCJ for ranking work, as it provided more structured comparisons and reduced subjectivity, but traditional marking remained valued for its transparency and feedback. The key takeaway was that MBCJ had strong potential as an alternative assessment method, particularly if mechanisms for providing direct feedback were integrated. Participants also suggested enhancements such as flagging close comparisons, incorporating an ``unsure" option, and using multiple markers to improve consistency.

    \subsection{Workshop Results and Analysis}

        At the start of the workshop, when the participants were asked if they felt that the distribution of the samples was evenly distributed between the three sub-samples. To which they all agreed they were.

        The workshop began with a recap of the three marking methods: Traditional Marking, BCJ, and MBCJ. Participants were invited to reflect on their initial assumptions about these methods before reviewing their marking outcomes. Most expected traditional marking to be the most transparent, given its structured, criteria-based approach and the ability to provide direct feedback to students. However, some had concerns about subjectivity and inconsistency, particularly when marking large cohorts. BCJ and MBCJ were seen as less familiar, and there was some scepticism about their fairness and accuracy compared to traditional methods.

        The discussion then moved to participants' experiences using the three marking methods. Traditional marking was perceived as the most cognitively demanding, requiring close reading of student work and continuous reference to marking criteria. Some found that they had to adjust their marks midway through once they had seen more submissions, leading to concerns about inconsistency. They also found that their early impressions of a submission could influence later judgements, particularly if a paper had poor presentation or grammatical errors. Despite these challenges, traditional marking allowed for detailed feedback, which participants felt was crucial for student improvement.
        
        BCJ was generally seen as less mentally taxing than traditional marking, as it focused on pairwise comparisons rather than absolute grading. However, some found the lack of explicit marking criteria challenging, as they were required to compare two pieces of work holistically rather than evaluating specific aspects. Participants noted that distinguishing between two similarly strong or weak submissions was difficult and some expressed concerns that the method felt more subjective than traditional marking. Transparency was also an issue—while rankings were generated, there was little insight into why one piece of work was ranked higher than another.
        
        MBCJ was viewed more favourably than BCJ, as it allowed participants to compare submissions based on multiple LOs rather than as a whole. This approach aligned better with their existing marking habits, as most participants were accustomed to grading based on specific assessment criteria. The ability to evaluate separate components of student work was seen as a major advantage over BCJ, reducing the feeling of subjectivity. Participants also noted that MBCJ was more structured and systematic, making it easier to justify ranking decisions compared to BCJ.
        
        After reviewing the ranking outcomes for each method, participants were surprised by the results. Traditional marking had the highest level of inconsistency, with tau scores revealing significant variation between markers. In contrast, BCJ and MBCJ produced rankings that were more consistent and closer to the target rankings. While some had initially believed that traditional marking would be the most accurate, the results suggested otherwise. The relative consistency of BCJ and MBCJ rankings challenged assumptions about the reliability of conventional assessment methods.
        
        The discussion then turned to trust and transparency. Initially, traditional marking was considered the most transparent because it provided explicit scores and justification for each mark. However, after seeing the ranking results, participants questioned whether transparency alone was enough if the method produced inconsistent outcomes. While BCJ and MBCJ lacked direct feedback, they were more reliable in producing fair rankings, which some participants argued could enhance trust in the system. A key challenge remained: how to integrate meaningful feedback into CJ methods.
        
        One of the major concerns was that BCJ and MBCJ, despite their improved consistency, did not provide students with detailed feedback on how to improve. Some participants suggested that automated feedback tools could be developed to provide comments based on ranking decisions. Others proposed a hybrid approach, where BCJ or MBCJ could be used for initial ranking, followed by targeted traditional marking for feedback. This could reduce cognitive load while maintaining transparency and student guidance.
        
        Participants also reflected on how marking scales over larger cohorts. Traditional marking was seen as impractical for large groups, as it required significant time and effort to maintain consistency across multiple markers. They discussed how CJ could help mitigate marker bias and inconsistency, particularly if multiple assessors were involved in ranking submissions. MBCJ was seen as particularly useful for moderation purposes, as it allowed different markers to contribute to a more reliable overall ranking. It was perceived that both BCJ and MBCJ would be most effective if it was that multiple markers were working on a larger pool of assessments together. Believing that and inconsistencies would then be corrected by the BCJ system's ranking abilities. Which is an interesting point as in usual CJ implementations, this is how CJ is usually carried out, as it can be down with one or multiple markers contributing together \cite{10.1145/3551708.3556204}. However, this approach was not implemented in this study. 
        
        By the end of the workshop, participants had significantly revised their views. Initially, most had assumed that traditional marking was the most trustworthy and accurate method, but the ranking results demonstrated that MBCJ was more consistent and less prone to bias. While BCJ was still viewed as somewhat subjective, MBCJ’s structured, multi-criteria comparisons made it a strong alternative to traditional marking. The main limitation remained the lack of direct feedback, which participants felt must be addressed before it could fully replace conventional methods.
        
        The workshop concluded with a discussion on future improvements. Participants suggested that flagging difficult comparisons, incorporating an “unsure” option, and integrating structured feedback tools could make CJ more effective. They agreed that, while traditional marking may remain necessary for providing feedback, MBCJ offered a more scalable, fair, and reliable method for ranking student work. The key takeaway was that MBCJ had the potential to replace traditional marking in many contexts, provided feedback mechanisms were developed.

    \subsection{Expert Interviews Results and Discussions}

        Three experts who research the CJ approach within assessment were interviewed for this section. Two of the experts interviewed work within government educational institutions, with one having previously worked for a UK exam awarding body while researching and implementing CJ, the third was an academic who researches CJ while also implementing it within their teaching practice. The experts were asked questions, in one-to-one semi-structured interviews (see Appendix \ref{app:expert_int}).

        Expert One discussed their current use of CJ, primarily for setting grade boundaries rather than direct marking. They noted that CJ is valuable for maintaining transparency and consistency because it focuses expert judgements on comparative quality rather than absolute scores. This reduces biases linked to knowing marks or year-specific expectations. However, they expressed caution about fully replacing traditional marking, as CJ's holistic nature introduces new biases, such as influences from handwriting or skipped questions, which do not affect absolute marking. Their organisation uses in-house tools for CJ experiments, allowing precise control over the exposure of the person doing the marking to submissions. They also employ rank-ordering for efficiency, though it sacrifices some intuitive usability.
    
        On transparency and reliability, the expert highlighted mixed views. From an educator's perspective, CJ is generally well-accepted and trusted for essay-based subjects but raises concerns about feedback limitations for students. The educators worry about inconsistent application of criteria across judges. The expert emphasised that CJ’s high reliability is not solely due to the comparative nature but rather, its multiple-judgement approach, collecting diverse opinions to minimise individual subjectivity. However, they cautioned against overstating CJ’s superiority, arguing that its reliability mirrors traditional marking when multiple markers are involved. They stressed that CJ is effective when multiple judges participate and question its value when only one judge is involved.
        
        Regarding BCJ, before seeing the proposed system, the expert noted assumptions that using a Bayesian model for estimating ranks does not inherently enhance transparency, as most users do not understand the underlying algorithms, which is the same for standard CJ. They argued that BCJ is as opaque as other statistical models used in CJ, especially when users are unaware of adjustments and estimation methods. They expressed concerns about potential ethical issues if priors were used, cautioning against bias in high-stakes assessments. However, they acknowledged that priors could accelerate processing in formative assessments, where fairness is less critical. They ultimately concluded that BCJ maintains the same transparency as standard CJ but with a more sophisticated estimation process.
        
        The expert found MBCJ to be promising in regard to improving transparency. They appreciated that MBCJ breaks down judgements by LOs, allowing judges to compare submissions across multiple criteria. This approach mirrors traditional marking more closely, helping educators justify rankings by explicitly assessing different dimensions of quality. They believed this reduces the holistic subjectivity found in standard BCJ, making it easier for judges to articulate their reasoning. The expert suggested that MBCJ could significantly improve transparency, especially if integrated with familiar reliability statistics to facilitate comparisons with traditional methods.
        
        In discussing transparency challenges, the expert pointed out that the holistic nature of the CJ inherently reduces transparency because judges make comparative decisions without justifying their reasoning. They acknowledged that training and monitoring judge fit can partially mitigate this, but transparency declines when using fewer criteria or discarding detailed mark schemes. They noted that MBCJ mitigates this issue by assessing LOs separately, but the absence of direct feedback still impacts transparency. They argued that feedback integration is essential for student understanding and trust in CJ systems.
        
        The interview ended with a discussion on future directions and tool development. The expert was enthusiastic about open-source tools to standardise CJ methodologies, reducing fragmentation in CJ research and practice. They recommended aligning MBCJ metrics with existing CJ reliability statistics, enabling more straightforward comparisons with traditional methods. They also highlighted the need for further research on feedback mechanisms, particularly how MBCJ could incorporate detailed feedback while maintaining its comparative strengths. Overall, they viewed MBCJ as a significant advancement, but stressed the importance of improvements in transparency, reliability, and usability to enable widespread adoption.

        Expert Two actively incorporates CJ into both research and teaching, with these applications sometimes overlapping but mostly remaining distinct. Their research focuses on CJ’s use in mathematics education, particularly to evaluate problem solving and conceptual understanding. Introduced to CJ in 2009, they have since expanded their work to comparing exam standards and exploring its use in philosophy, English literature, and psychology. In teaching, they have used CJ for peer-assessment, particularly with undergraduate and foundation mathematics students, and their early research in 2014 investigated its use in calculus peer assessment. They noted that CJ is engaging and practical for peer assessment, reducing the need to recruit external judges.

        They find CJ to be a reliable assessment method, especially with expert judges, though student-led peer assessments produce slightly lower reliability scores, which can be mitigated by increasing the number of judgements. While CJ enhances student engagement and motivation compared to traditional assessment, it faces several challenges. Institutional resistance is a key issue, as many universities require clear assessment criteria, often interpreted as rubrics and mark schemes, making CJ difficult to implement where policies mandate traditional approaches. Another challenge is feedback, as CJ does not naturally provide direct, personalised feedback, which is expected in many assessment frameworks. Transparency is also a concern, with CJ lacking a clear audit trail compared to rubrics, though Expert Two argues that all assessment methods have some degree of opacity and that CJ simply makes these issues more visible. They also highlighted the technological barriers, as early implementation required third-party platforms like \emph{No More Marking}, though they have now developed a Moodle plugin to facilitate CJ within their institution. Finally, they acknowledged that any educational innovation requires upfront effort and that CJ, like any new approach, relies on early adopters to gain traction.

        When discussing transparency and reliability, they found that CJ consistently delivers high reliability in assessment outcomes. While students as judges produce slightly lower reliability scores than experts, increasing the number of judgements compensates for this. In terms of transparency, they believe CJ is no more or less transparent than traditional assessments, but because it does not produce a breakdown of marks like a rubric, it can be perceived as opaque. However, their students do not seem to experience this issue, likely due to the structured way CJ is introduced and embedded into the learning process.

        Before seeing the results, Expert Two assumed that BCJ would likely face the same transparency issues as standard CJ. While it could provide a reliable ranking, they expected it to still be perceived as opaque, especially without explicit marking criteria. They thought that MBCJ might improve transparency by making the decision process clearer, but they were unsure whether it would fully resolve concerns about the lack of an audit trail. They also admitted to having limited familiarity with the underlying mathematical models but expressed enthusiasm for more sophisticated approaches to CJ, viewing advancements like BCJ and MBCJ as necessary for both research and educational assessment.

        When comparing marking approaches, it was discussed that the markers felt that traditional marking was seen as trustworthy and transparent, whereas BCJ was considered trustworthy but opaque, with some markers reporting cognitive strain when making close comparisons. MBCJ was generally preferred, by the markers, over BCJ, as it aligned more naturally with traditional marking practices and was perceived as easier to use. The results showed that MBCJ performed similarly to BCJ in terms of reliability but was favoured by markers due to reduced cognitive load and greater clarity in decision-making. After seeing these results, markers shifted their preference strongly towards MBCJ, as it combined the perceived trustworthiness of traditional marking with the efficiency and consistency of CJ. This shift away from traditional marking and toward MBCJ intrigued the expert.

        Expert Two found MBCJ particularly promising for educational assessment, especially in structured exam marking, and acknowledged the growing interest in multidimensional assessment models in research. However, they expressed concerns about potential interdependencies between criteria and stated that they would not use MBCJ for research purposes, as they prefer maintaining full independence between assessment criteria. They concluded that assessment should embrace multiple methods rather than enforcing a single standard and saw CJ, BCJ, and MBCJ as valuable tools that should be further explored depending on context and purpose.
    
        Expert Three provided a detailed account of their experience with CJ, highlighting its evolution and key challenges. While they no longer use CJ extensively in their current role, their early work focused on standard maintenance rather than using CJ as an alternative to traditional marking. Their initial experiences involved manually comparing physical scripts to link standards between different exams, such as A-level Maths syllabi from different decades. This process was slow and logistically complex, prompting them to explore ranking multiple items instead of making only pairwise comparisons. However, they noted that analysis still required converting rankings back into pairs, which could sometimes create a misleading impression of reliability.

        Reflecting on transparency and reliability in CJ, Expert Three acknowledged that the process is often seen as a “black box,” making it difficult to trace how final rankings and marks are determined. While CJ leverages human ability to make relative judgements naturally, the transformation of those judgements into final scores is less intuitive. They expressed concerns about fairness in adaptive CJ models, where different scripts might not be judged against the same set of comparisons. Ensuring reliability was another key issue, particularly in research focused on expert bias and the extent to which judges properly account for differences in test difficulty. They emphasised that public understanding of CJ outcomes depends on clear communication, noting the importance of developing a compelling narrative to support its credibility.
        
        Discussing BCJ and MBCJ, Expert Three found the approaches intriguing but had questions about how the distributions were modelled and how prior knowledge evolved over successive comparisons. They were particularly interested in MBCJ, noting that it could improve transparency by allowing judges to evaluate multiple assessment criteria separately. This could align more closely with traditional assessment practices, where different attributes of student work—such as structure, engagement, and argumentation—are typically considered independently.
        
        When reflecting on the presentation and interpretation of results, Expert Three found the visualisations generated by BCJ and MBCJ to be useful but noted that they would require significant training to be fully understood. They raised concerns about the scalability of these methods, questioning whether they could function effectively in large-scale assessments such as GCSEs, where thousands of scripts must be processed. While acknowledging that BCJ and MBCJ worked well in small-scale university settings, they suggested that implementing them on an industrial level would introduce new challenges. Nevertheless, they saw potential in MBCJ for addressing validity concerns, particularly in preventing snap judgements based purely on first impressions.
        
        Looking ahead, Expert Three suggested further research into the generalisability of BCJ and MBCJ in large-scale assessments. They also highlighted ongoing work in the field, including studies on how judges make decisions and process information when using CJ. One significant challenge they identified was the difficulty of providing meaningful feedback to students within a CJ framework, an issue that remains a key area for development. Despite these challenges, they expressed keen interest in the continued evolution of BCJ and MBCJ and requested access to research papers detailing these approaches.

        The expert interviews highlighted both the strengths and limitations of CJ-based methods. While CJ was valued for its consistency, concerns about transparency persisted, particularly in explaining how rankings are derived and interpreted. Its holistic nature allowed for flexibility, but made it difficult to justify individual decisions, presenting challenges for both assessors and students.

        BCJ was seen as a refined extension of CJ, offering a structured statistical approach but raising concerns about transparency due to its complex calculations. Ethical considerations around priors in high-stakes settings were noted, though they were viewed as useful in formative assessments. MBCJ, however, was received more favourably as it aligned with traditional marking by assessing multiple LOs separately. This distinction improved transparency by allowing judges to justify rankings based on clear criteria rather than an overall impression.
        
        A key theme was the shift in perception after reviewing BCJ and MBCJ results. Initially, traditional marking was preferred for its perceived transparency, while BCJ was seen as reliable but opaque. However, MBCJ emerged as the preferred approach, maintaining high reliability while reducing cognitive strain and providing clearer decision-making structures. Experts noted that this change in preference underscored the importance of usability and training in the success of new assessment models.

    \subsection{BCJ Transparacy in the Assessment Procedure}
        Initially, traditional marking was perceived as the most transparent because it provided a structured, criteria-based approach with explicit marks and feedback. Participants felt that transparency came from clearly defined assessment rubrics, where each score was justified based on LOs. However, they also acknowledged that traditional marking relied heavily on the individual marker’s judgement, which could introduce inconsistencies between assessors. Some participants noted that transparency was undermined by subjectivity, as different markers might interpret criteria differently, particularly in open-ended or qualitative assessments.

\begin{figure}[ht!]
                \centering
                \includegraphics[width=\linewidth]{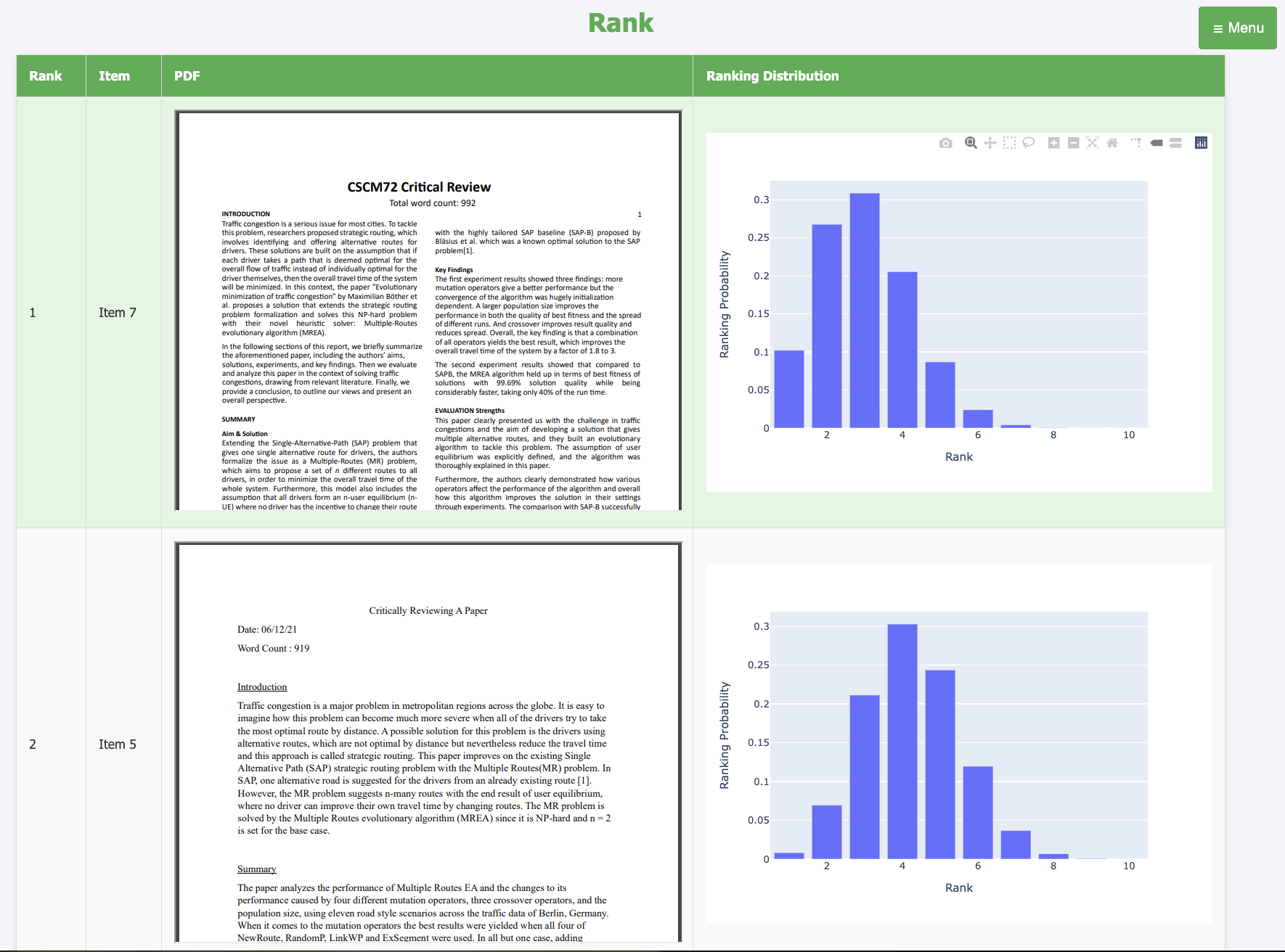}
                \caption{This figure shows the results probability distribution and the ranking after the pair-wise comparisons for BCJ.}
                \label{fig:bcj_transparancy}
            \end{figure}

        BCJ was widely seen as less transparent than traditional marking, as it lacked explicit justifications for ranking decisions. Participants found it difficult to determine why one submission was ranked higher than another, as the process was holistic and comparative rather than criteria-based. The ranking system felt somewhat like a ``black box", where the final order of submissions emerged without a clear rationale for individual placements (see Figure  \ref{fig:bcj_transparancy}), apart from the transparency that the systems produce in displaying the ranking probabilities. This lack of insight made some participants feel less confident in the fairness of BCJ, even though the method produced more consistent rankings than traditional marking.

\begin{figure}
                \centering
                \includegraphics[width=\linewidth]{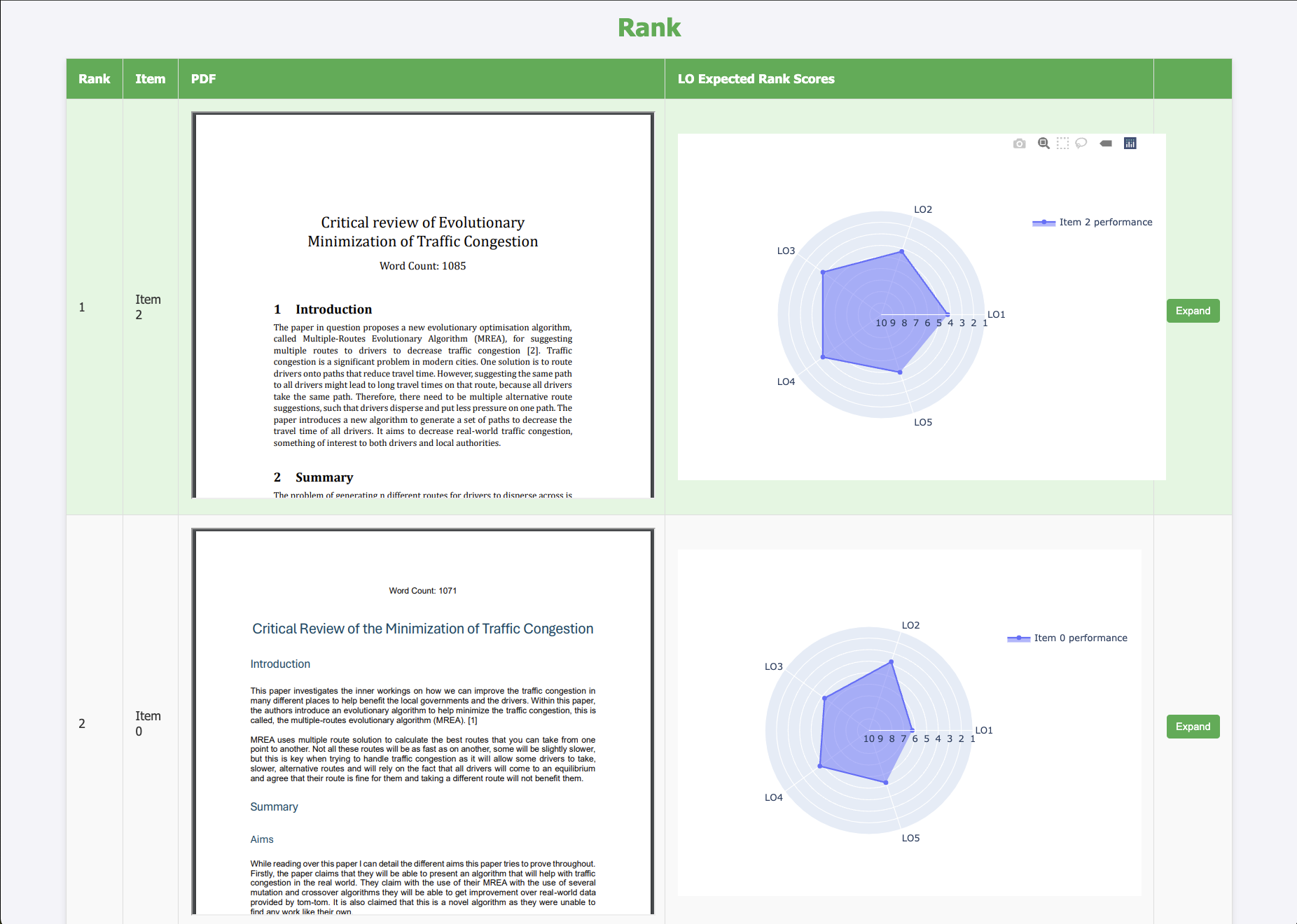}
                \caption{This figure shows the radar plots, showing where an item scored against each criterion, and the ranking for each item after the pair-wise comparisons for BCJ.}
                \label{fig:transparancy1}
            \end{figure}

\begin{figure}
                \centering
                \includegraphics[width=\linewidth]{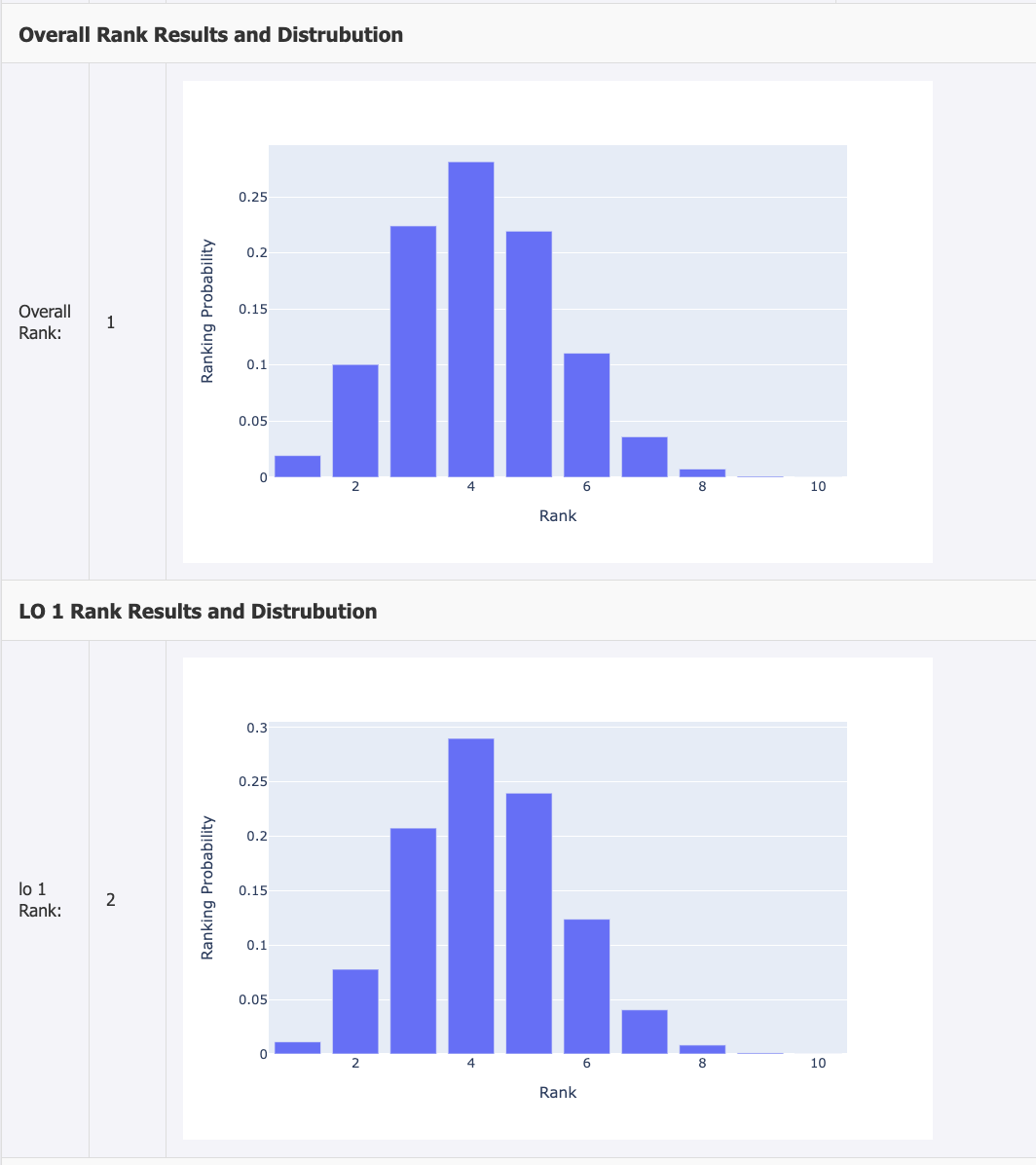}
                \caption{This figure shows the the results and ranking distribution to an item's overall ranking and individual criterion ranking. This information is displayed when the expand button has been pressed.}
                \label{fig:transparancy2}
            \end{figure}

        MBCJ, however, was seen as a step towards greater transparency (see Figures \ref{fig:transparancy1}, \ref{fig:transparancy2})
        . Since it broke down comparisons across multiple LOs, participants felt that the ranking process was more structured and aligned with how they naturally assessed student work. Unlike BCJ, MBCJ provided clearer insights into why one submission was stronger in specific areas, which made it easier to justify ranking decisions. While MBCJ did not provide direct explanations for individual scores, its structured nature reduced the perception of randomness in the process, making it feel more transparent than BCJ.
        
        During the workshop, participants' views on transparency in traditional marking changed after seeing the results. While they initially viewed traditional marking as the most transparent, the ranking inconsistencies between markers raised concerns about its reliability. The high variability in scores across assessors suggested that transparency did not necessarily equate to fairness, as two different markers could produce significantly different results despite following the same marking scheme. This led some participants to reconsider whether transparency in process outweighed the need for reliability in outcomes.
        
        A major issue discussed was the role of feedback in transparency. Traditional marking was still preferred in terms of transparency because it allowed markers to explicitly communicate reasoning to students. In contrast, BCJ and MBCJ, despite being more consistent in ranking, did not naturally provide detailed feedback on areas for improvement. Participants felt that without feedback, transparency was limited, as students would not fully understand why they received a particular ranking or how they could improve. This was seen as a critical barrier to adopting CJ methods in student assessments.
        
        Participants agreed that transparency must be balanced with reliability and fairness. While traditional marking was still valued for clear justification and student feedback, its inconsistencies reduced trust in the process. BCJ was viewed as too opaque for individual assessments, but MBCJ provided a reasonable compromise by offering structured rankings across criteria. The consensus was that MBCJ had the potential to be a transparent and fair alternative to traditional marking, but only if mechanisms for providing student feedback were integrated into the process.

                

\begin{figure*}
                \centering
                \includegraphics[width=\linewidth]{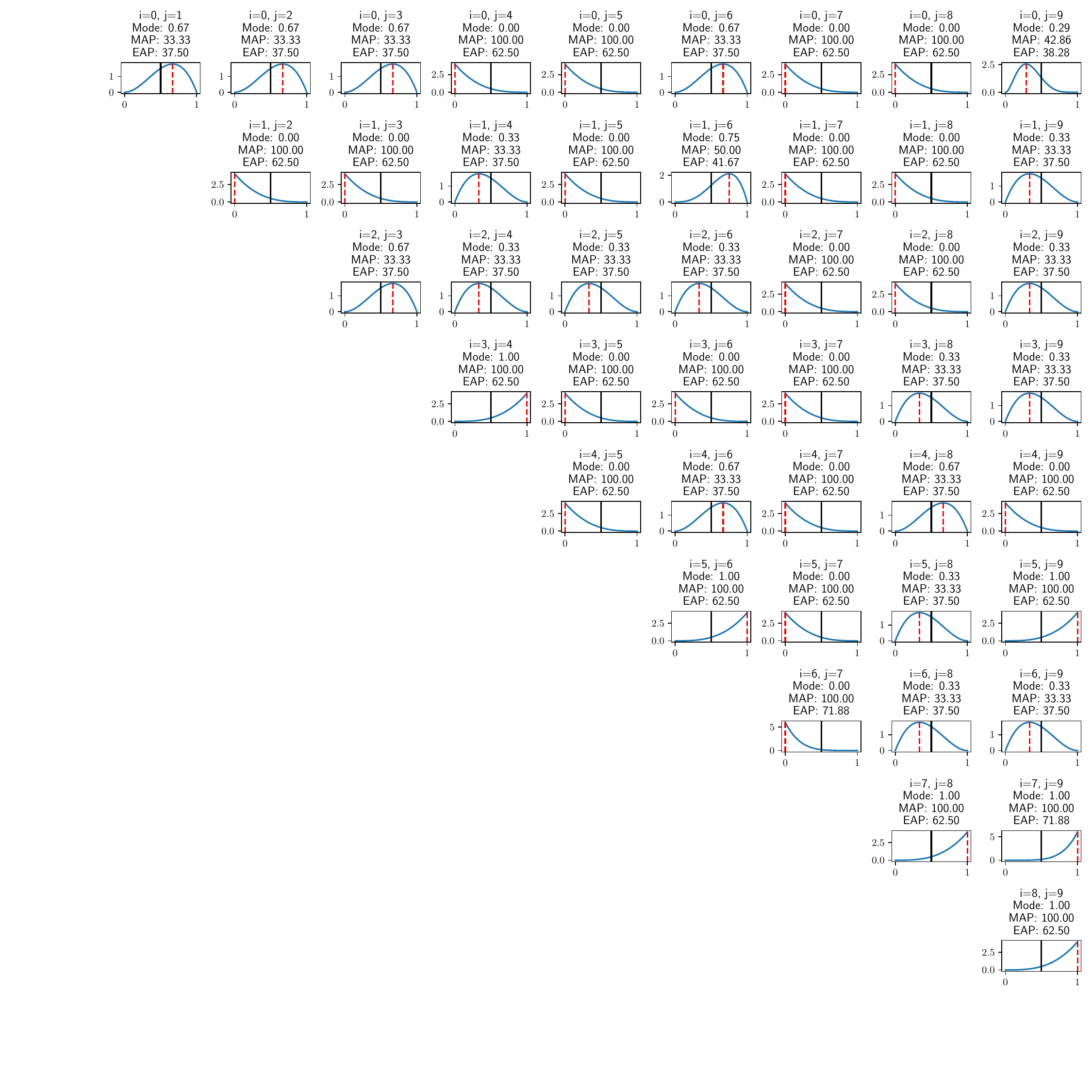}
                \caption{This figure shows the transparency in the decisions being made. The closer the distribution is to $0.5$ (the black line), the more uncertain the markers are, meaning that half went one way and the other half went the other way. The red dotted line represents the mode $\beta$ value from the decisions made, while the blue line shows the probability density function of the $\beta$ distribution.}
                \label{fig:decision_transparancy}
            \end{figure*}

        By using the BCJ or MBCJ approach, we can also display the transparency in the decisions being made, identifying where a maker or marker has diverged from each other (see Figure \ref{fig:decision_transparancy}). If the markers agree, the distribution and the mode score will be close to $1$ or $0$, depending on whether they prefer item a or b. However, if the mode is close to $0.5$, then this shows uncertainty in the preferences shown by the markers. With $0.5$ indicating that $50\%$ went for item a, while $50\%$ went for item b, as being the better item of the two. Ultimately, this provides even more detailed transparency on the results.

\begin{figure}
            \centering
            \begin{subfigure}{\linewidth}
                \centering
                \includegraphics[width=\textwidth]{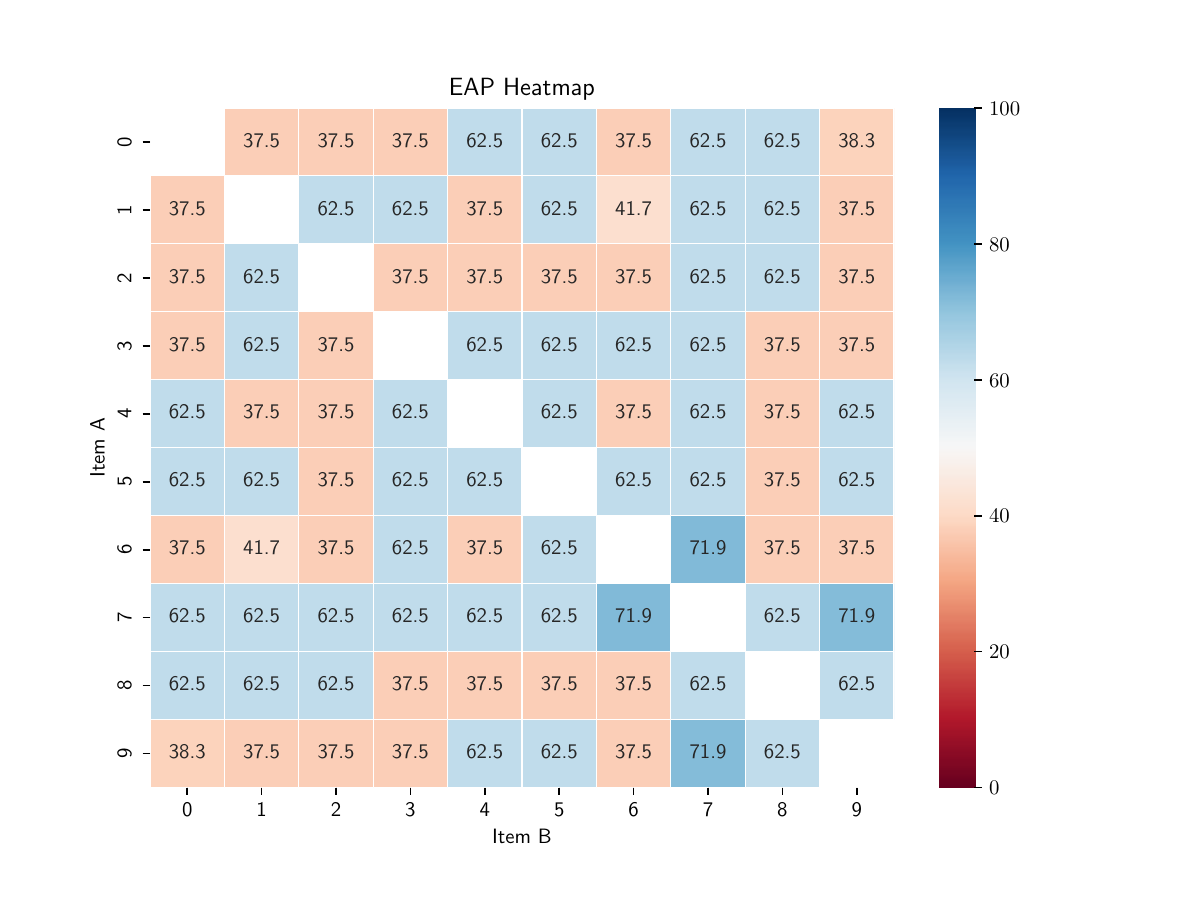} 
            \end{subfigure}
            \hfill
            \begin{subfigure}{\linewidth}
                \centering
                \includegraphics[width=\textwidth]{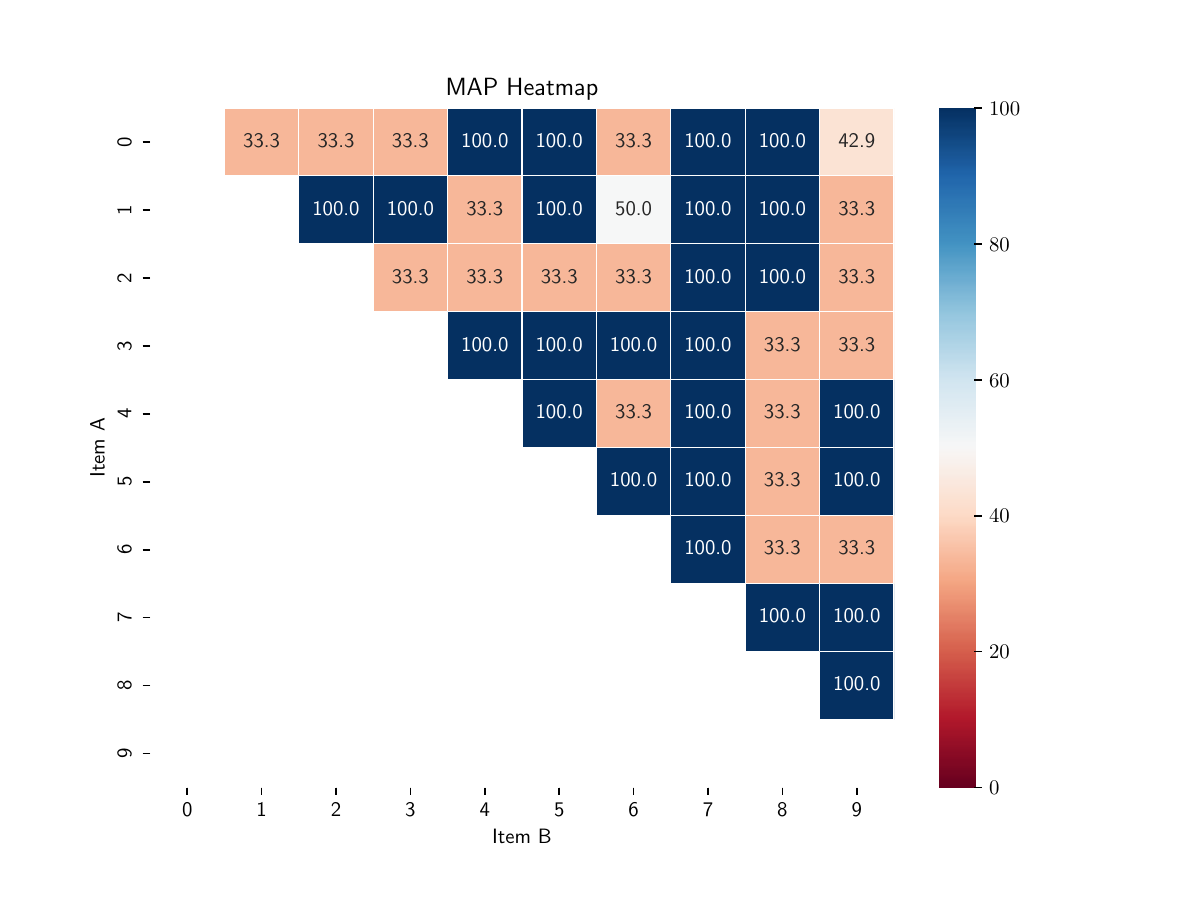} 
            \end{subfigure}
            \caption{The figure shows an example of the EAP and MAP outputs. These heatmaps can be produced for all LOs for the MBCJ and holistically for the BCJ approach. Any value $\geq$ to $50$ shows that the agreement is outside the $25^{th}$ and $75^{th}$ percent quantile ranges.}
            \label{fig:MAP_EAP}
        \end{figure}

        With the mode agreement percentage (MAP) and the expected agreement percentage (EAP) proposed by \cite{gray2025bayesianactivelearningmulticriteria}, we can additionally create a heatmap that produces these scores (see Figure \ref{fig:MAP_EAP}). These heatmaps show a value between one and zero, with a score $\geq 0.5$ representing a good score. If the values of MAP and EAP are equal or above $0.5$, then that means the markers are mostly in agreement and that their decisions are outside of the $0.25$ - $0.75$ quartile range, showing a preference for one item over the other. This is what we would expect if all markers agree that one item is better than another, as displayed in figure \ref{fig:decision_transparancy}, where $i=0$ and $j=5$. While any value less than $0.5$ indicates that the judgements are within the $0.25$ - $0.75$ quartile range, identifying that there is some disagreement between the markers, with the closer to $0$ these scores achieve the closer the PDF and mode displayed in figure \ref{fig:decision_transparancy} is to $0.5$ which means that there is a complete disagreement between the markers, representing that $50\%$ went one way and $50\%$ went the other way.

        Expert two, when shown these outputs (Figures \ref{fig:decision_transparancy}, \ref{fig:MAP_EAP}), found the insights informative and suggested that these metrics could be a good alternative to measuring reliability compared to the current approach of Scale Separation Reliability (SSR).


    \subsection{Implementing BCJ}

        While the web apps used within this experiment are open-sourced and available on GitHub (see section \ref{sec:ack} for links), the documentation has been provided to make the process as seamless as possible. However, at the point of writing, there are elements of the web app that the users will have to adapt to use for themselves manually. These changes are explained within the GitHub repository's README file. However, while no great deal of coding knowledge is required, having coding experience will undoubtedly help with the process.

        Additionally, for large sample sizes, the ranking process can be resource-heavy. Therefore, depending on the specifications of the machines being used, the ranking process can take some time. Still, the process for both standard and MBCJ for comparing and deciding on the next pair to present to the assessor is relatively quick.



\section{Conclusions}
    \label{sec:con}

    BCJ has shown to be an effective tool for assessing students' work. Both the standard and multi-criteria approach to BCJ as a tool for enhancing transparency in educational assessment. BCJ directly enhances transparency for students and assessors by how the model makes decision-making more visible and understandable by providing the ranking distributions to the assessors and the students, therefore enabling greater transparency in assessments.

    While both methods provided enhanced transparency, the multi-criteria approach provided greater in-depth transparency on how the students performed across each individual LO they were being assessed on, as well as the holistic overall performance. This enabled a greater understanding of where students rank compared to their peers. However, the multi-criteria approach did involve additional thought to be made per item and its individual LO; the greater transparency the approach offered is a worthy trade-off. However, if educators didn't want the detailed response of the multi-criteria approach, the standard approach of BCJ was well received by the assessors. 

    This study has explored the potential of BCJ and MBCJ as alternatives to traditional assessment methods. By applying these approaches in a real-world educational setting, we have demonstrated how they can enhance transparency, consistency, and fairness in student ranking.


    The findings highlight that while traditional marking remains the most familiar and widely trusted approach, it is also the most cognitively demanding and prone to inconsistencies between markers. BCJ and MBCJ, in contrast, offer a structured way to compare student work, reducing the subjectivity that can arise in absolute marking. The results indicate that BCJ improves reliability by systematically tracking prior judgements, while MBCJ further enhances this process by allowing assessments to be broken down by specific LOs. This provides a clearer picture of student strengths and weaknesses, making the assessment process more transparent and informative.
    
    Educators participating in the study found that MBCJ, in particular, aligns more closely with their existing marking practices, as it preserves the depth of feedback associated with rubric-based assessment while offering the efficiency and fairness of CJ. However, a key limitation identified was the need for mechanisms to provide detailed feedback to students, ensuring they understand not only their ranking but also how they can improve.
    
    Moving forward, further research should explore how these methods can be integrated into broader assessment frameworks, including their potential for high-stakes examinations and large-scale implementation. Additionally, developing feedback mechanisms within BCJ and MBCJ could help bridge the gap between ranking-based assessment and student learning.
    
    Therefore, this study supports the idea that structured CJ methods have the potential to improve transparency and fairness in assessment, providing educators with a data-informed approach that maintains educational integrity while reducing workload pressures. 
    However, further features to the process is required in order to make it have more adoption within a classroom setting.

\section*{Author contributions}


    Andy Gray: Writing – review \& editing, Writing – original draft, Software, Project administration, Methodology, Investigation, Formal analysis, Data curation, Conceptualisation. Alma Rahat: Writing – review \& editing, Conceptualisation, Supervision. Tom Crick: Writing – review \& editing, Supervision. Stephen Lindsay: Writing – review \& editing, Supervision. Jen Pearson: Writing – review \& editing, Supervision.

\section*{Acknowledgments}
    \label{sec:ack}
    Andy Gray 
    is funded by the EPSRC Centre for Doctoral Training in {\emph{Enhancing Human Interactions and Collaborations with Data and Intelligence-Driven Systems}} (EP/S021892/1) at 
    Swansea University. 
    Additionally, the project stakeholder is 
    CDSM. 
    We are particularly grateful to their CIO, 
    Darren Wallace. 
    For the purpose of Open Access, the author has applied a CC-BY public copyright licence to any Author Accepted Manuscript (AAM) version arising from this submission.
    All underlying data to support the conclusions are provided within this paper.

    Ethics approval for the use of the secondary data was approved by the Faculty of Science and Engineering ethics committee at Swansea University (Research Ethics Approval Number: 1 2023 7465 6926).

    The code for the single-dimension app can be found here: \url{https://github.com/codingWithAndy/BayesCJ-Web-App}.
    The code for the multi-criteria app can be found here: \url{https://github.com/codingWithAndy/BayesCJ-multi-dimensional-Web-App}.

\bibliographystyle{elsarticle-num} 
\bibliography{cas-refs}

\bio{}
\endbio


\appendix

\section{Questionnaire}
    \label{app:HCI_questions}

    \section*{Introduction}
    We are conducting this survey to understand your experiences with three different marking methods: Traditional Marking, Bayesian Comparative Judgement, and Multi-Criteria Bayesian Comparative Judgement. Your feedback is valuable and will help us improve these methods. This survey should take about 10-15 minutes to complete.

    \section*{Experience with Marking Methods}
        
        For each of the following sections, please rate your experience on a scale of 1 (Very Poor) to 5 (Excellent).

        Definitions: 
        \begin{itemize}
            \item Transparent: The clarity and openness of the processes, criteria, and decisions involved in evaluating students’ work. It ensures that students, teachers, and other stakeholders understand how judgements are made, promoting fairness and trust in the outcomes.
            \item Traditional absolute marking: Evaluating students' work against a fixed set of criteria or a predetermined mark scheme, where grades are awarded based on the extent to which these criteria are met.
            \item Bayesian Comparative Judgement (BCJ): Assessing students' work by comparing pairs of pieces and using a probabilistic model to estimate the quality of each, based on the collective judgements made.
            \item Multi-criteria BCJ: Multi-criteria comparative judgement involves comparing pairs of students' work across multiple criteria or dimensions, allowing for a holistic evaluation that simultaneously considers various aspects of quality.
        \end{itemize}

    \section*{Traditional Marking}
        \begin{enumerate}
            \item How would you rate the ease of use of Traditional Marking (1-5) and why?
            \item How transparent do you find the process of Traditional Marking (1-5) and why?
        	\item How accurate do you find the results of Traditional Marking?
        	\item What were your initial impressions of the marking approach?
            \item What do you think is happening when you use traditional marking methods?
            \item Do you enjoy using this approach to mark the student's work?
            \item What alterations would you want to see to it?
            \item Does this approach to marking work fit with how you think about student work?
        \end{enumerate}
     
    \section*{Bayesian Comparative Judgement}
        \begin{enumerate}
        \setcounter{enumi}{8}
            \item How would you rate the ease of use of Bayesian Comparative Judgement (1-5) and why?
            \item How transparent do you find the process of Bayesian Comparative Judgement (1-5) and why?
            \item How confident are you in the ranks derived from the Bayesian Comparative Judgement (1-5) and why?
            \item What were your initial impressions of the marking approach?
            \item What do you think is happening when you use Bayesian Comparative Judgement?
            \item Would you recommend this approach over traditional marking? Why?
            \item What alterations would you want to see to it?
            \item Does this approach to marking work fit with how you think about student work?
        \end{enumerate}

    \section*{Multi-criteria Bayesian Comparative Judgement}

    \begin{enumerate}
    \setcounter{enumi}{16}
        \item How would you rate the ease of use of the multi-criteria Bayesian Comparative Judgement (1-5) and why?
        \item How transparent do you find the process of the multi-criteria Bayesian Comparative Judgement (1-5) and why?
        \item How confident are you in the ranks derived by using the multi-criteria Bayesian Comparative Judgement (1-5) and why?
        \item What were your initial impressions of the marking approach?
        \item What do you think is happening when you use Multi-criteria Bayesian Comparative Judgement?
        \item Would you recommend this approach over traditional marking? Why?
        \item What alterations would you want to see to it?
        \item Does this approach to marking work fit with how you think about student work?
    \end{enumerate}

    \section*{Additional Feedback}
    \begin{enumerate}
    \setcounter{enumi}{24}
        \item Which marking method did you prefer and why?
        \item Which approach do you have the most confidence in their rankings of the work and Why?
    \end{enumerate}
    
    Thank you for your time and feedback!

\section{Workshop}
    \label{app:workshop}

    \section*{1. Welcome and Introduction 
    }
\begin{itemize}
    \item Recap the three marking methods: Traditional Marking, Bayesian Comparative Judgement (BCJ), and Multi-criteria Bayesian Comparative Judgement (MBCJ).
    \item Outline the objectives of the workshop:
    \begin{itemize}
        \item Explore participants' experiences with the marking methods.
        \item Present marking outcomes, including metrics such as tau scores and rank comparisons.
        \item Discuss the implications of these outcomes on trust and transparency.
        \item Evaluate transparency and trustworthiness from a student perspective.
    \end{itemize}
\end{itemize}

\section*{2. Assumptions and Processes 
}
\begin{itemize}
    \item Facilitate an exploration of participants' assumptions about each method:
    \begin{itemize}
        \item What assumptions did you make about the marking processes?
        \item Did the results or tau scores challenge these assumptions?
        \item How do the results reflect fairness or accuracy in marking?
    \end{itemize}
    \item Draw connections between assumptions, outcomes, and trust in the methods.
\end{itemize}

\section*{3. Group Reflection: Experiences with Marking Methods 
}
\begin{itemize}
    \item Divide participants into small groups to discuss:
    \begin{itemize}
        \item How does your experience with the marking methods align with your performance outcomes?
        \item Which method do you feel most confident using, and why?
        \item Did the performance comparisons (tau scores) change your perceptions of the methods?
    \end{itemize}
    \item Summarise group discussions and share key insights.
\end{itemize}

\section*{4. Presentation of Marking Outcomes 
}
\begin{itemize}
    \item Present participants' marking outcomes:
    \begin{itemize}
        \item \textbf{Performance Metrics:} Show how participants' rankings align with the target rank using the tau score.
        \item Compare individual ranks against each other.
    \end{itemize}
    \item Facilitate discussion:
    \begin{itemize}
        \item What do these metrics reveal about the reliability of each method?
        \item How do the results align with your expectations?
        \item Does seeing the comparative performance affect your trust in the methods?
    \end{itemize}
\end{itemize}

\section*{5. Trustworthiness and Transparency Discussion 
}
\begin{itemize}
    \item Facilitate an open discussion using prompts:
    \begin{itemize}
        \item Which method do you trust the most after seeing the results? Why?
        \item From a student perspective:
        \begin{itemize}
            \item Which method best conveys fairness and transparency?
            \item Would knowing about tau scores or ranking comparisons build or undermine student trust?
        \end{itemize}
    \end{itemize}
\end{itemize}

\section*{6. Reflection and Actionable Feedback 
}
\begin{itemize}
    \item Ask participants to reflect individually on:
    \begin{itemize}
        \item Key insights about the strengths and weaknesses of each method.
        \item Suggestions to improve usability, trust, or transparency.
        \item How seeing their performance outcomes impacts their overall views.
    \end{itemize}
    \item Collect feedback using sticky notes or a shared digital board.
\end{itemize}

\section*{7. Closing and Next Steps 
}
\begin{itemize}
    \item Thank participants and recap key points discussed.
    \item Highlight how their feedback will shape future developments in marking practices.
\end{itemize}

\section{Expert Semi Structured Interview}
    \label{app:expert_int}

\section*{Phase 1}

\subsection*{First Element - Your CJ Practice}
\begin{itemize}
    \item Overview of your current CJ practice.
    \item Key methodologies and approaches used.
    \item Challenges faced in implementing CJ.
    \item Effectiveness of CJ in assessment.
    \item Potential improvements and refinements.
\end{itemize}

\subsection*{Second Element - Transparency and Reliability of CJ}
All elements are considered from:
\begin{itemize}
    \item Student perspective
    \item Educator perspective
\end{itemize}

Key aspects of transparency and reliability:
\begin{itemize}
    \item Transparency of the mark derived – how rank informs the mark and how rank is decided.
    \item How student work informs the rank that is derived – the elements of it that influence it.
    \item Reliability of the markers doing the work and following the rules.
    \item Transparency of the comparison selection process (random vs. Bayesian).
\end{itemize}

\subsection*{Third Element - BCJ and MBCJ for Transparency}
\begin{itemize}
    \item Explanation of BCJ and its role in assessment.
    \item How BCJ enhances transparency in the judgement process.
    \item Introduction to Multi-Criteria BCJ and its advantages.
    \item Comparison of BCJ and Multi-Criteria BCJ in terms of reliability and fairness.
    \item Challenges and potential improvements in implementing BCJ-based systems.
\end{itemize}

\vspace{5mm}

\section*{Phase 2}

\subsection*{Fourth Element - Presentation and Reflection on Results}
Key considerations:
\begin{itemize}
    \item Are the results convincing?
    \item Are the results comprehensible?
    \item Do they address the questions that the expert has about MBCJ?
\end{itemize}

\subsection*{Fifth Element - Future Design}

What further data could be gathered to support the MBCJ concept?

Achieving transparency outside of research settings if Tau scores cannot be used:
\begin{itemize}
    \item Seeded judgments where the lead knows the answer.
    \item Identifying outlier marking.
\end{itemize}

\end{document}